\documentclass[lettersize,journal]{IEEEtran}

\usepackage{colortbl}
 \usepackage[table,xcdraw]{xcolor}
\usepackage{graphicx} % Required for inserting images

\usepackage{latexsym}
\usepackage{amssymb}
\usepackage{amsmath}
\usepackage{amsthm}
\usepackage{booktabs}
\usepackage{enumitem}
\usepackage{graphicx}
\usepackage{color}
\usepackage{multirow}
\usepackage{tikz}
\usepackage{pifont}
\usepackage{subfig}
\usepackage{hyperref}

\usepackage{tikz}
\usepackage{calc}
\usepackage{pgfplots}
\usepackage{pgfplotstable} % For better data handling in tables
\pgfplotsset{compat=1.18}

\def\checkmark{\tikz\fill[scale=0.4](0,.35) -- (.25,0) -- (1,.7) -- (.25,.15) -- cycle;} 
\def\scalecheck{\resizebox{\widthof{\checkmark}*\ratio{\widthof{x}}{\widthof{\normalsize x}}}{!}{\checkmark}}

\def\checkmark{\tikz\fill[scale=0.4](0,.35) -- (.25,0) -- (1,.7) -- (.25,.15) -- cycle;}

\title{\textit{Detection Made Easy}: Potentials of Large Language Models for Solidity Vulnerabilities}
\author{
  \IEEEauthorblockN{
    Md Tauseef Alam,
    Raju Halder,
    and
    Abyayananda Maiti
  }
\thanks{Md Tauseef Alam, Raju Halder, and Abyayananda Maiti are with the Department of Computer Science and Engineering, Indian Institute of Technology Patna, India, 801106. (Email: \{tauseef\_2121cs04,halder,abyaym\}@iitp.ac.in)}
\thanks{This work has been submitted to the IEEE for possible publication. Copyright may be transferred without notice, after which this version may no longer be accessible.}
}

\begin{document}

\maketitle

\begin{abstract}
The large-scale deployment of Solidity smart contracts on the Ethereum mainnet has increasingly attracted financially-motivated attackers in recent years. A few now-infamous attacks in Ethereum's history includes DAO attack in 2016 (\$50 million lost), Parity Wallet hack in 2017 (\$146 million locked), Beautychain’s token BEC in 2018 (\$900 million market value fell to 0), and NFT gaming blockchain breach in 2022 (\$600 million stolen). This paper presents a comprehensive investigation of the use of large language models (LLMs) and their capabilities in detecting OWASP Top Ten vulnerabilities in Solidity. We introduce a novel, class-balanced, structured, and labeled dataset named VulSmart, which we use to benchmark and compare the performance of open-source LLMs such as CodeLlama, Llama2, CodeT5 and Falcon, alongside closed-source models like GPT-3.5 Turbo and GPT-4o Mini. Our proposed SmartVD framework is rigorously tested against these models through extensive automated and manual evaluations, utilizing BLEU and ROUGE metrics to assess the effectiveness of vulnerability detection in smart contracts. We also explore three distinct prompting strategies—zero-shot, few-shot, and chain-of-thought—to evaluate the multi-class classification and generative capabilities of the SmartVD. Our findings reveal that SmartVD outperforms its open-source counterparts and even exceeds the performance of closed-source base models like GPT-3.5 and GPT-4o Mini. After fine-tuning, the closed-source models, GPT-3.5 Turbo and GPT-4o Mini, achieved remarkable performance with 99\% accuracy in detecting vulnerabilities, 94\% in identifying their types, and 98\% in determining severity. Notably, SmartVD performs best with the `chain-of-thought' prompting technique, whereas the fine-tuned closed-source models excel with the `zero-shot' prompting.
\end{abstract}

%Large-scale deployment of Solidity smart contracts on Ethereum mainnet opens an avenue for many financially-motivated attackers in recent years. Few now-infamous attacks that took place in the history of Ethereum includes DAO hack in 2016 leading to a loss of \$150 USD. 

\section{Introduction} \label{sec:Introduction}

\IEEEPARstart{T}{he proliferation} of blockchain platforms, such as Ethereum, in recent years has driven the widespread adoption of smart contracts across various sectors \cite{aggarwal2019blockchain}. However, with this popularity and mass-scale adoption, security vulnerabilities in Solidity smart contracts have surfaced, causing substantial economic losses \cite{hewa2021survey}. These security challenges not only put users' funds at risk but also impede the overall growth of the  blockchain ecosystem. As a result, identifying and addressing vulnerabilities in smart contracts becomes increasingly crucial.

Large language models (LLMs) have recently demonstrated impressive capabilities in a wide range of language-related tasks \cite{chang2024survey}. LLMs undergo pre-training with extensive text corpora, utilizing various self-supervised training strategies, including masked language modeling and next sentence prediction \cite{chang2024survey}. LLMs such as GPT-4 and CodeLlama, which are pre-trained on extensive code corpora, have achieved significant progress in various code-related tasks within a relatively short time. These tasks encompass code completion \cite{husein2024large}, automated program repair \cite{zhang2024systematic}, test generation \cite{wang2024software}, code evolution, and fault localization \cite{kang2024quantitative}, among others. The success of these models underscores the potential of Large Language Models (LLMs) and opens the door to exploring more advanced techniques. This leads to an interesting question: \emph{Can these cutting-edge pre-trained LLMs also be effectively applied to detecting security vulnerabilities in Solidity smart contracts?}

In this context, several existing works \cite{chen2023chatgpt,napoli2023evaluating,boi2024vulnhunt,xia2024auditgpt,10431564,sun2024gptscan,10.1145/3687251.3687253} investigated the use of large language models (LLMs), such as GPT models (e.g., GPT-3, GPT-3.5 Turbo), for detecting Solidity vulnerabilities.  For instance, \cite{chen2023chatgpt} evaluated ChatGPT's vulnerability detection capabilities using SmartBugs-curated benchmark codes \cite{10.1145/3324884.3415298}. However, as the dataset exclusively contains only vulnerable contracts, this potentially led to misleading results due to hallucinations caused by the imbalance in the dataset. Even though, the authors in \cite{10431564, sun2024gptscan, 10.1145/3687251.3687253} assessed the performance of LLMs, such as Llama2 and GPT models, on vulnerability detection, they provided only preliminary insights with limited in-depth analysis. While other studies \cite{napoli2023evaluating, boi2024vulnhunt, xia2024auditgpt} utilized ChatGPT for vulnerability detection and remediation, their primary emphasis was on fixing vulnerable codes. %The works in \cite{10431564, sun2024gptscan} provide an assessment of GPT's capabilities in smart contract auditing, offering preliminary insights with limited in-depth analysis.

%Furthermore, datasets like SmartBugs-curated \cite{10.1145/3324884.3415298}, which include vulnerability information in the form of comments, can bias the LLMs under consideration. 

These initial efforts fail to give a clear picture of what impact is made by fine-tuning process needed for correct knowledge enrichment. They also lack to give an idea of open source LLMs and closed source LLMs to proceed in this domain. Importantly, these works often overlook the critical role of prompt engineering techniques in enhancing vulnerability detection. Therefore, there is a pressing need to move beyond these initial efforts and comprehensively evaluate the potential of state-of-the-art LLMs in smart contract vulnerability detection. Additionally, the exploration of open-source LLMs in this domain remains limited, warranting further investigation. 

While vulnerability detection is inherently complex with numerous factors that can influence and potentially limit the quality of solutions, this can significantly be improved with the advanced expressiveness and broad knowledge incorporated into state-of-the-art large language models (LLMs). Effective vulnerability detection requires a thorough and sophisticated understanding of different elements within the code, highlighting its importance as a key research area. Investigating this hypothesis is crucial for progressing in the field of vulnerability detection. Initial results, as presented in \cite{chen2023chatgpt, david2023you}, indicate that simple prompting techniques with the ChatGPT LLM did not yield significantly better outcomes compared to random prediction in the context of vulnerability detection. This insight underscores the value of specifically fine-tuning large language models (LLMs) for this task, suggesting it as a promising avenue for future research exploration.
%One such factor may be the limitations in the capacity and code comprehension abilities of models like CodeLlama. 

This paper presents \textit{SmartVD}, an efficient LLM-driven vulnerability detection framework for Solidity codes, which is developed by fine-tuning the pre-trained LLM, Codelama. To facilitate this fine-tuning, we introduce a carefully curated domain-specific class-balanced dataset, called \textit{VulSmart}. Our approach involves two key methods to enable a detailed and accurate assessment by the model: a) binary classification, which determines whether vulnerabilities exist within functions, and b) multi-class classification and generation, which identifies specific types of vulnerabilities along with its severity. This evaluative methodology is consistently applied to both traditional deep learning classification techniques and other popular LLMs under review, enabling us to perform a rigorous assessment on the effectiveness of \textit{SmartVD} in comparison to other pre-trained LLMs. We also attempt to understand how fine-tuning the LLMs on our labeled datasets affects the model's performance in this classification task. Additionally, we conduct a manual review and alteration of selected code examples to uncover which features LLMs can and cannot recognize when predicting vulnerabilities. Finally, we incorporate semantic structures into the LLMs to assess their impact on the models' effectiveness in detecting vulnerabilities.
% \todo{add line for dataset and source code}

%To achieve this goal, a comprehensive and balanced dataset is essential. Existing datasets in the literature \cite{9072659, crytic, 10.1145/3395363.3397385, 10.1145/3324884.3415298, 10.1145/3460319.3464837, yashavant2022scrawlddatasetrealworld} are typically unstructured and therefore not directly suitable for training deep learning-based vulnerability detection tools, highlighting the need for a more structured dataset. Additionally, two major challenges—labeling and class imbalance—persist, which we address in our work. We introduce a class-balanced dataset, \textit{VulSmart} (\textit{Vul}nerability detection for \textit{Smart} Contracts), detailed in Section \ref{corpus}.

%developed by fine-tuning the pre-trained LLM, Codelama, with a well-defined labeled rich domain-specific dataset, called \textit{VulSmart}, prepared by us. Our comprehensive investigation evaluates the effectiveness of \textit{SmartVD} compared to the pretrained Large Language Models (LLMs). 
%Our comprehensive investigation assesses the effectiveness of \textit{SmartVD} in comparison to other pre-trained Large Language Models (LLMs). 
%In particular, we conduct a thorough evaluation to determine whether LLMs can effectively classify a Solidity code snippet as either containing or not containing a particular type of vulnerability. 

\subsection{Research Questions}
We formulate the following research questions to meet our aforementioned objectives. \newline
\textit{\textbf{RQ1.} How do various pre-trained LLMs and traditional deep learning techniques perform in detecting security vulnerabilities of Solidity smart contracts?
\newline
\textbf{RQ2.} How does fine-tuning of pre-trained LLMs using our proposed domain-specific dataset, \textbf{VulSmart}, impact the performance of vulnerability detection in smart contracts?\newline
%
%How do different pre-trained LLMs perform in detecting security vulnerabilities of smart contracts after fine-tuning on \textbf{VulSmart} dataset?
%
\textbf{RQ3.} How do different prompting techniques influence the performance of both pre-trained and fine-tuned LLMs in vulnerability detection?\newline
%
%How do different prompting techniques impact the efficacy of pre-trained large language models (LLMs) as well as fine-tuned LLMs?
%
\textbf{RQ4:} To what extent do adversarial attacks affect the vulnerability detection capabilities of fine-tuned LLMs? \newline
%
%Can adversarial attacks impact the performance of fine-tuned LLMs in detecting vulnerabilities? 
%
\textbf{RQ5:} Can the integration of semantic structures, such as control-flow and data dependencies, enhance the performance of LLMs in identifying vulnerabilities?
}
% \newline\textbf{RQ5:} Is our proposed method useful for detecting
% new types of vulnerabilities, e.g., sharing-variable
% reentrancy, which is difficult for existing methods?

\subsection{Contributions}
To sum up, our key contributions are as follows:

\begin{itemize}
    \item To the best of our knowledge, we conduct the first large-scale study to quantitatively and qualitatively measure the performance of open source and closed LLMs in the field of smart contracts vulnerability detection. In addition, we conduct a comparative study of deep learning classification frameworks and language models.

    \item We develop a class balanced dataset, \textit{VulSmart}, for \textit{Vul}nerability detection for \textit{Smart} Contracts.

    \item We propose a multi-class classification  and generation framework \textit{SmartVD} for vulnerability detection that harnesses the unique capabilities of LLMs to generate type of vulnerability and its severity of a vulnerable smart contract within the specified dataset.

    \item We validate our framework by assessing various prompting techniques utilized in LLMs, intending to determine the most efficient approach for addressing vulnerability related concerns.

\item  Tapping into the unexplored potential of LLMs in vulnerability detection, our results provide valuable insights and pave the way for future research in this field.

\item We plan to make datasets, source codes and models accessible for further research at https://anonymous.4open.science/r/DetectionMadeEasy/.
\end{itemize}

\section{Related Studies}
% here we can use the dataset, fm and DL thing from paper "A survey on smart contract vulnerabilities: Data sources, detection and repair" also refer "SurveySCTools-July23"

% class imbalance point file:///Users/mdtauseefalam/Downloads/electronics-13-02295.pdf 

The following work have been relevant to the following research areas, namely smart contract vulnerability detection using: a) Formal method tools, and b) Large Language Models (LLMs)

\subsection{Formal Method Tools}

In the work \cite{ivanov2023security} Nikolay Ivanov et al. highlight 133 proposals published in field of smart contract vulnerability out of which 115 solutions are based on formal method techniques. Out of these 115 tools, as highlighted in \cite{wei2023comparative} because of non-availability, non-functionality, non-compatibility and no documentation we only have few popular working tools. Oyente \cite{luu2016making} is the pioneer work for smart contract bug detection based on symbolic execution from solidity bytecode. However, it suffers from several false alarms even in case of trivial Smart Contracts (for Reentrancy). Moreover, verification conditions adopted in this tool are neither sound nor complete. Slither \cite{feist2019slither} is an open source static analysis tool developed by trail of bits. It uses intermediate representation SlithIR (SSA based) for solidity code and predefined rules to match the problematic codes. Since, it uses predefined rules to identify the vulnerability it tends to give several false positives.
 Krupp et al. develop teEther \cite{krupp2018teether}, a symbolic execution based tool to create an end to end exploit generation from smart contracts bytecode. It constructs control flow graph from bytecode to generate path constraints, solved by theorem prover Z3. However, these unreadable bytecodes make detection inefficient and rules difficult to formulate. Tsankov et al. develop Securify \cite{tsankov2018securify} at SRI labs (ETH Zurich). It works at bytecode level and derives semantic facts from contract's dependency analysis. Then use this information to check compliance and violation patterns for proving if a property written in domain specific language holds or not. Smartcheck \cite{smartcheck} translates solidity source code to XML-based intermediate representation checked against XPath patterns to find vulnerability. However, the intermediate representation restrictions make it difficult to formulate vulnerability rules. Mythril \cite{mythril} employs symbolic execution to build the Control Flow Graph (CFG) of the contract from the EVM bytecode, and then executes the predefined logic rules to identify vulnerabilities. Manticore \cite{mossberg2019manticore} uses dynamic symbolic execution technique to detect bug for traditional binaries and Ethereum bytecode. This tool suffers from number of contracts checking failed due to predefined assumptions. Wang et al. proposes SliSE \cite{wang2024efficiently} that uses program slicing and symbolic execution technique to detect reentrancy vulnerability among inter contracts. However, it also uses the fixed pattern to analyse the bug which leads to false positives.

\subsection{Large Language Models}

Chen et al. \cite{chen2023chatgpt} explore the potential of ChatGPT in detecting smart contract vulnerability types. They compare ChatGPT with other formal method tools and identify the cause of false positives generated by ChatGPT. In \cite{napoli2023evaluating} the authors evaluate the performance of ChatGPT in fixing vulnerable code. They check for the vulnerability in existing code and then pass the error message along with code to fix it. The authors in \cite{boi2024vulnhunt} propose an approach VulnHunt-GPT using GPT3 to detect smart contract vulnerability by training it on smart contract functions and vulnerability. They also compare their results with the existing formal method tools. In \cite{xia2024auditgpt} Xia et al. uses the large language model to verify the ERC specification against its solidity source code implementation to better understand its security impact. The authors in \cite{10.1145/3687251.3687253} leverage the large language model Llama2 to detect the vulnerability present in smart contracts more effectively. They train the LLM on custom dataset to demonstrate its efficacy in terms of accuracy when compared to formal method based tools. In \cite{10431564} Hu et al. propose GPTLens framework where the LLM plays the dual role of Auditor and Critic. Auditor finds the vulnerability and critic validates it and thus reduces the false positives. In \cite{soud2024soley} the authors introduce Soley an automated approach using LLMs to detect logic vulnerabilities in smart contracts. In \cite{sun2024gptscan} the authors employ static analysis combined with GPT-3.5 Turbo to detect vulnerabilities in smart contract logic. They introduce GPTScan, a tool that utilizes GPT to identify potential vulnerabilities, enhancing accuracy by guiding GPT to intelligently recognize key variables and statements. These identified elements are then validated through static confirmation.

% \todo{Include GPTScan[
% https://daoyuan14.github.io/papers/ICSE24_GPTScan.pdf
% ],  }

% Evaluating ChatGPT, vulnHuntGPT, AuditGPT, Boi et. al. [https://dl.acm.org/doi/abs/10.1145/3687251.3687253], hu et. al. [https://ieeexplore.ieee.org/abstract/document/10431564] auditor-critic,

% soley tool [https://arxiv.org/pdf/2406.16244]

% \todo{A comparative table highlighting the existing works drawbacks tackled in our work}

% \textit{\textbf{Limitations:} As evident from literature that there have been initial works on using LLMs to detect vulnerabilities in smart contracts. However, these lacks to give a clear picture of what impact is made by fine-tuning process needed for correct knowledge enrichment. They also lacks to give an idea of open source LLMs and closed source LLMs to proceed in this domain. These also lack to show the impact of prompt engineering for vulnerability detection. We have addressed these limitations along with performing an extensive large-scale study on LLMs in this domain.
% }

\subsection{Existing datasets}

The existing popular dataset in the field of smart contracts reported in literature are CodeSmell \cite{9072659} which has smart contract related posts from Ethereum Stack Exchange, as well as 587 real-world smart contracts containing defects. SolidiFi \cite{10.1145/3395363.3397385} contains 300 real world smart contracts injected with seven different vulnerability types. SmartBugs (curated and wild) \cite{10.1145/3324884.3415298} consists of 143 annotated vulnerable contracts with 208 tagged vulnerabilities and 47,518 unique contracts collected through Etherscan. Not So Smart Contract \cite{crytic} consists of 18 real world smart contracts containing bug which lead to substantial economic loss, Smart Contract benchmark Suites \cite{10.1145/3460319.3464837} consists of 393 real world vulnerable smart contracts compiled from different github repository and sources. ScrawlID \cite{yashavant2022scrawlddatasetrealworld} consists of 6780 smart contracts address and the vulnerability present. As evident, all these existing datasets are imbalanced and lack localized information regarding the vulnerabilities present within the code.

\begin{table*}[!hbtp]
\caption{A comparative summary of existing datasets.}
\label{tab:tableDataset}
\begin{tabular}{cccccccc}
\hline
\multicolumn{1}{c|}{\textbf{Sl. No.}} & \multicolumn{1}{c|}{\textbf{Datasets}} & \multicolumn{1}{c|}{\textbf{\begin{tabular}[c]{@{}c@{}}\# Smart\\ Contracts\end{tabular}}} & \multicolumn{1}{c|}{\textbf{Class Balanced}} & \multicolumn{1}{c|}{\textbf{\begin{tabular}[c]{@{}c@{}}\# Types of \\ Vulnerability\end{tabular}}} & \multicolumn{1}{c|}{\textbf{Structured}} & \multicolumn{1}{c|}{\textbf{Labelled}} & \textbf{Source Code} \\ \hline
1. & CodeSmell \cite{9072659} & 587 & $\times$ & 20 & \scalecheck & \scalecheck & $\times$ \\
2. & Solidifi \cite{10.1145/3395363.3397385} & 350 & $\times$ & 7 & $\times$ & \scalecheck & \scalecheck \\
3. & SmartBugs-curated \cite{10.1145/3324884.3415298} & 143 & $\times$ & 9 & $\times$ & \scalecheck & \scalecheck \\
4. & SmartBugs-wild \cite{10.1145/3324884.3415298} & 47,398 & $\times$ & 0 & $\times$ & $\times$ & \scalecheck \\
5. & Not So Smart Contract \cite{crytic} & 18 & $\times$ & 12 & $\times$ & \scalecheck & \scalecheck \\
6. & Smart Contract Benchmark Suites \cite{10.1145/3460319.3464837} & 46,186 & $\times$ & 8 & $\times$ & \scalecheck\parbox[t]{0pt}{\footnotesize\textsuperscript{1}} & \scalecheck \\
7. & ScrawlID \cite{yashavant2022scrawlddatasetrealworld} & 6,780 & $\times$ & 8 & $\times$ & \scalecheck & $\times$ \\ \hline
\textbf{8.} & \textbf{VulSmart [Ours]} & \textbf{6,085} & \scalecheck & \textbf{13} & \scalecheck & \scalecheck & \scalecheck \\ \hline
\end{tabular}
\begin{tabbing}
    \textsuperscript{1} \scriptsize{214 smart contracts are annotated with labels.}
\end{tabbing}
\end{table*}

\section{Corpus Description} \label{corpus}

 Given the critical importance of smart contract security, accurate data on vulnerabilities is essential to identify real-world contracts that are at risk, along with the specific type of vulnerabilities they contain. Therefore, a comprehensive dataset that includes both smart contract source code and a detailed and localized vulnerability information is crucial before applying it to LLMs. Our study began with an extensive literature review to assess existing datasets related to smart contract vulnerabilities. However, we found that most existing datasets \cite{9072659, crytic, 10.1145/3395363.3397385, 10.1145/3324884.3415298, 10.1145/3460319.3464837, yashavant2022scrawlddatasetrealworld} suffer from a significant lack of structured, labeled data, and many are imbalanced as evident from Table \ref{tab:tableDataset}, making them unsuitable for effective training or fine-tuning of LLMs.

To the best of our knowledge, no publicly available dataset offers a class balanced collection of smart contract code, along with corresponding vulnerability types and localized information. To address this gap and improve smart contract security, we have curated a novel dataset, \textit{VulSmart}, which includes the source code of smart contracts, the type of vulnerability, its severity, and the specific vulnerable code segments. Table \ref{statistic} provides the statistics of this balanced dataset, which consists of 1,125 samples encompassing 6,085 smart contracts. In the following sections, we outline the steps taken to develop this dataset.
 
% https://openreview.net/pdf?id=Q3GVrWRKuB#page=11&zoom=100,110,316 Table 6:
% \todo{The types of vulnerabilities contained in each dataset.}

% The existing datasets lack diversity. In the domain of smart contract vulnerability detection, the supply of labeled datasets is notably scarce. Compounding this issue, the vulnerability fragments produced by current generation techniques are often strikingly similar, even though there may be a vast array of fragment pools. 

% Machine learning-based
% methods for vulnerability detection either create their new dataset or exploit one of the following
% ones available.

% Although Ethereum is a public blockchain
% with millions of blocks, only a small fraction of stored contracts have accessible source code. This
% is because only the bytecode is required to be stored on the blockchain, not the source code.
% A contract’s code is accessible only if the developer explicitly includes it during deployment. Furthermore, many of these contracts are identical, which significantly reduces the number of
% unique contracts available. In this context, Smartbugs Wild represents the most viable alternative,
% being the largest dataset of unique Ethereum smart contracts available, with about 47,000 contracts.
% Despite this extensive collection, two major issues persist: labeling and class imbalance. Significant
% efforts are still needed to address these issues. Currently, a combination of manual labeling and static
% analyzers represents the state-of-the-art approach, with careful consideration of their limitations.

\subsection{Dataset Collection}

All smart contracts deployed on the Ethereum network can be accessed through Etherscan. However, only a small portion of these contracts have publicly available source codes, as bytecode is the only requirement for deployment on the network, leading many developers to withhold their source code. Additionally, there are numerous identical smart contracts with available source code, reducing the number of unique contracts. This makes the task of collecting smart contract source code particularly challenging.

With this in mind, we adopt the principle of re-usability and select the dataset corpus hosted on GitHub by various sources, containing source code from real-world smart contracts that have been featured in peer-reviewed publications. This way, we ensure a broad and diverse representation within the dataset. We collect a total of 9,437 smart contracts source code in this phase. These contracts include both vulnerable and non-vulnerable examples, verified using formal methods tools. Notably, the vulnerability information in existing dataset is often included as comments within the source code, making it challenging to use directly for our objectives. As a result, we proceeded to the next step of data cleaning.

\subsection{Data Cleaning}

We conduct data cleaning by processing the collected data from the previous step. The dataset of 9.4K collected smart contracts consists of both vulnerable and non-vulnerable contracts, with approximately 78\% classified as non-vulnerable. To create a class-balanced dataset, we carefully selected 3,000 samples from this pool. The selection is based on appropriate contract size, with incomplete, non-functional, or corrupted contracts excluded through verification using the Remix IDE\footnote{https://remix.ethereum.org}. To eliminate duplicates, we first remove comments and extraneous blank lines from the source code. We then use the hash of the source code to further ensure no duplicate samples remain. Afterward, we proceed to data annotation, classifying each smart contract as either vulnerable or non-vulnerable, ultimately creating a balanced dataset with both types of samples.

% We performed data cleaning by processing the collected
% data from the previous step. The dataset contained a mix of vulnerable and non vulnerable contracts. From this pool, 3000 samples were selected meticulously filtered on the basis of reasonable contract size and removing incomplete or corrupted contracts. We then remove the comments and additional blank lines from source code. We do this to remove the duplicates in the sample. We then use the hash of source code to remove the duplicates samples. We then move to the next phase of data annotation by checking the smart contract as vulnerable/non vulnerable and create an balanced dataset with both type of samples.
% 2. Remove comments and additional new lines from the smart contracts

\subsection{Data Annotation}

% can include few things from data set creation of the following ppr [https://arxiv.org/pdf/2406.16244]

The previous phase results in a corpus of 2,541 unique samples, consisting of smart contract source code along with their respective compiler versions. We adhere to the \textit{OWASP Smart Contract Top 10}\footnote{https://owasp.org/www-project-smart-contract-top-10/} and \textit{DASP top 10}\footnote{https://dasp.co/} guidelines to target the most critical vulnerabilities out of the 37 listed in the Smart Contract Weakness Classification (SWC) registry\footnote{https://swcregistry.io/}. For vulnerability severity classification, we follow the Smart Contract Security Verification Standard (SCSVS)\footnote{https://github.com/ComposableSecurity/SCSVS} and utilize SmartCheck \cite{smartcheck}.

% The above phase yielded a corpus comprising a set of unique 2,541 samples containing source code of smart contracts with the compiler versions. We followed the guidelines of \textit{OWASP smart contract top 10}\footnote{https://owasp.org/www-project-smart-contract-top-10/} and \textit{DASP top 10}\footnote{https://dasp.co/} to select and target the most notorious vulnerabilities out of 37 listed in Smart Contract Weakness Classification (SWC) registry\footnote{https://swcregistry.io/} and for the severity classification of the type of vulnerability we followed Smart Contract Security Verification Standard (SCSVS)\footnote{https://github.com/ComposableSecurity/SCSVS} and SmartCheck \cite{smartcheck}. 

In total, we target 13 types of vulnerabilities, including Access Control, Arithmetic Overflow/Underflow, Bad Randomness, Denial of Service, Front Running/Transaction Ordering Dependence, Gasless Send, Reentrancy, Short Addresses, Time Manipulation, tx.origin, Unchecked Low-Level Call, Unsafe Delegate Call, and Unsafe Suicide. These vulnerabilities are classified into three severity levels: High, Medium, and Low.

\begin{figure}[!htbp]
\centerline{\includegraphics[width=0.5\textwidth]{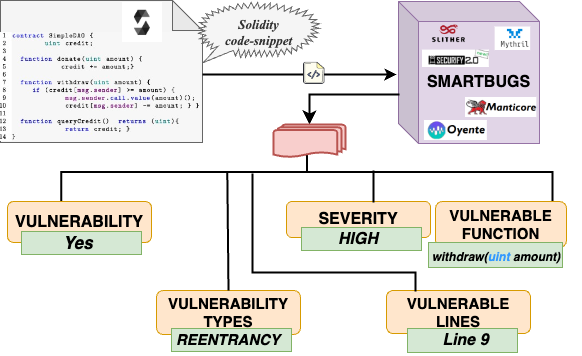}}
\caption{Flowchart depicting the annotation process for the \textit{`VulSmart'} dataset}
\label{dataAnnotate}
\end{figure}

We select four popular state-of-the-art (SOTA) formal method tools—Securify, Slither, Mythril, and Oyente—to detect the presence of these vulnerabilities. A vulnerability is annotated in a smart contract if at least two of the four tools identify the same issue. The annotation process of \textit{VulSmart} dataset is depicted in Figure \ref{dataAnnotate}. We first provide the source code to the Smartbugs framework \cite{10.1145/3324884.3415298}, which contains docker images of the four tools, and monitor the log files generated by each tool. After reports are generated, we label each smart contract sample based on the presence of vulnerabilities (Yes/No), the specific vulnerability type from the 13 under consideration, its severity (High/Medium/Low), the vulnerable function, and the corresponding vulnerable lines.

The entire annotation process took approximately one month, completed by two trained undergraduate students under the guidance of a senior PhD scholar specializing in formal methods. To ensure the quality of the annotations, we measure inter-annotator agreement (IAA) using Cohen's Kappa score \cite{viera2005understanding}, resulting in a score of 0.76, affirming the high quality and reliability of the annotations.

% \todo{here u can show the diagram of data annotation flow chart}

\subsection{Data validation}

After completing the annotations of the dataset samples, we proceed with manual data validation. For this task, we invite two postgraduate students and one PhD scholar from the computer science department, each with technical expertise in formal methods. Together, they validate 10\% of the dataset samples. Once validation was complete, the labeled contracts are added to finalize our \textit{VulSmart} dataset for further use.

% Once the annotations of the dataset samples is completed we manually go for data validation. we selected two post graduate students and one PhD scholar from computer science department having technical expertise in formal methods for data validation. We validated the 10\% of the samples in the dataset. Subsequently, we add the labeled contracts to finalized our \textit{VulSmart} dataset for further use.

Our objective is to offer researchers a foundational resource that can be further expanded with additional examples—ideally, encompassing different forms of the same vulnerability.

% Our objective is to offer researchers a fundamental resource that can be further developed by incorporating further examples, ideally encompassing various forms of the same vulnerability.

% Our goal is to provide a foundational resource for researchers, while also allowing the dataset to be expanded with additional examples—ideally, incorporating different forms of the same vulnerability.

% \todo{Write first about dataset used. SBcurated, SBWild....}
% In this study, we use the......

% The source of the dataset is twofold: ........ For data labeling, we utilize Mythril, Slither, Oyente, and SmartCheck to detect vulnerabilities in the collected contracts. When three out of four tools report issues with a specific line of the contract, we manually verify and label the specific type of vulnerability after validation. Subsequently, we add the labeled contracts to the vulnerability dataset.

% To address this limitation, we propose
% \textit{VulSmart}, the first multi-labeled dataset specific to smart contract vulnerability detection.

\begin{table}[!hbtp]
\centering
\caption{Statistics of {\em VulSmart} Dataset}
\label{statistic}
\begin{tabular}{ll}
\hline
\textbf{Measures} & \textbf{Size} \\ \hline
\textit{Number of Total Samples} & 1125 \\
\textit{Number of Smart Contracts in Total Samples} & 6085 \\
\textit{Number of Functions in Total Samples} & 26618 \\
\textit{Number of Lines of Code in Total Samples} & 293421 \\
\textit{Number of True labels (Vulnerability) in Total Samples} & 484 \\
\textit{Number of False labels (Vulnerability) in Total Samples} & 641 \\
\textit{Count of Types of Vulnerability in Total Samples} & 13 \\
\textit{Number of High labels (Severity) in Total Samples} & 384 \\
\textit{Number of Medium labels (Severity) in Total Samples} & 99 \\
\textit{Number of Low labels (Severity) in Total Samples} & 1 \\
\textit{Number of Not Mentioned labels (Severity) in Total Samples} & 641 \\ \hline
\end{tabular}
\end{table}
% \todo{Add types of vulnerability also}

\section{Proposed Methodology}

\subsection{Problem Definition}

% \section{Problem Definition}
Given a smart contract \( C \) written in Solidity, the objective is to identify and generate critical information regarding potential vulnerabilities within the code. These tasks are defined as follows:

\begin{enumerate}
    \item \textbf{Vulnerability Detection:}
    \begin{itemize}
        \item We define a characteristic function \( f_1: C \rightarrow \{0, 1\} \) such that:
        \[
        f_1(C) =
        \begin{cases}
        1 & \text{if } \exists \, v \in \mathcal{V}, \\ & \text{where}\ \mathcal{V} \text{ is the set of  vulnerabilities} \\
        0 & \text{otherwise}
        \end{cases}
        \]
        This function detects the existence of any vulnerability \( v \) in the contract \( C \).
    \end{itemize}
    
    \item \textbf{Vulnerability Type Identification:}
    \begin{itemize}
        \item Given that \( f_1(C) = 1 \), we introduce a mapping \( f_2: C \rightarrow \mathcal{T} \), where \( \mathcal{T} \) denotes the set of all possible vulnerability class. This mapping identifies the class/type \( T \in \mathcal{T} \) of the detected vulnerability.
    \end{itemize}
    
    \item \textbf{Severity Assessment:}
    \begin{itemize}
        \item We define a severity function \( f_3: \mathcal{T} \rightarrow \mathcal{S} \), where \( \mathcal{S} \) represents the set of all severity levels, structured as a partially ordered set (poset) \( (\mathcal{S}, \leq) \). The function \( f_3(T) = S \) evaluates the severity \( S \in \mathcal{S} \) corresponding to the identified vulnerability type \( T \).
    \end{itemize}
\end{enumerate}

\begin{figure}[!htbp]
\centerline{\includegraphics[width=0.5\textwidth]{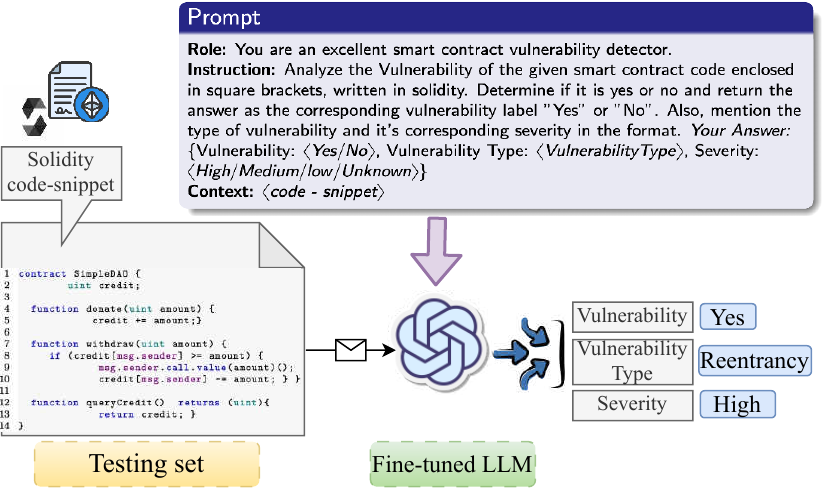}}
\caption{An instance of Multi-Class Classification}
\label{classi}
\end{figure}

The entire process can be described by a composite function \( F: C \rightarrow \mathcal{V} \times \mathcal{T} \times \mathcal{S} \), defined as
\(
F(C) = \left(f_1(C), f_2(C), f_3(f_2(C))\right)
\).
This composite function sequentially applies \( f_1(C) \), \( f_2(C) \), and \( f_3(T) \) to the smart contract code \( C \), generating a tuple \( (v, T, S) \) that encapsulates information about the presence (\( v \)), type (\( T \)), and severity (\( S \)) of the detected vulnerabilities. The output is a well-defined and structured data that characterizes the vulnerabilities present in the smart contract \( C \). Figure \ref{classi} illustrates an instance of multi-class classification applied to address the problem defined above. In this instance, the source code of a smart contract is provided to the LLM along with a prompt. The model then outputs the presence of a vulnerability (binary classification), its type (multi-class classification), and the corresponding severity level (multi-class classification).
% using formal mathematical constructs.

% \subsection{Multi-Class Classification}

% \textbf{Vulnerability Classification:}

% \textbf{Vulnerability Type Classification:}

% \textbf{Severity Classification}

% In addition, we conducted a comparative study of deep learning classification frameworks and language models.

\subsection{Selection of LLMs}

We highlight the challenges we encountered during implementing our methodology and selection of the LLMs.

\begin{itemize}
    \item Out of several LLMs, we identify and fine-tune four optimal code-related models. While these LLMs excel in code comprehension and generation, they lack inherent capabilities for vulnerability detection. The aim was to select models that are easy to use from a technical perspective (i.e. model size, available libraries, inference speed etc). For instance, CodeGeeX \cite{zheng2023codegeex} does not provide standard library support for training, which limited its applicability.

    \item We meticulously select large language models (LLMs) with configurations aimed at optimizing resource efficiency and minimizing computational overhead. In our selection process, we carefully consider both the context window—the maximum number of tokens the model can process as input—and the total number of tokens supported by the LLM. This approach is essential to streamline the development of our SmartVD framework, ensuring that it operates effectively without excessive resource demands.

    \item Existing pre-trained language models are typically designed for tasks like next token prediction, which differs from our objective of vulnerability classification. Unlike token prediction, our goal is to categorize the entire input sequence rather than anticipate the following tokens. To enhance the performance of LLMs for this specific task, we incorporate vulnerability-related features directly into the smart contract code during training. Despite these efforts, we observe that SolidityT5\footnote{https://huggingface.co/hululuzhu/solidity-t5} do not perform optimally for classification tasks.

\end{itemize}

\subsection{Components of SmartVD Framework}

Our \textit{SmartVD} framework is fine-tuned version of Code Llama \cite{codellama_paper} and the following are its components:

\textbf{1) Input Layer:} The input layer is composed of smart contract code snippets and the corresponding instructions provided as prompts.

\textbf{2) Tokenizer:} The smart contract code, provided as input text, undergoes tokenization using the AutoTokenizer class, where it is divided into a sequence of tokens. \(
t = \text{AutoTokenizer}(C)
\), where \( C \) is the input smart contract code and \( t \) is the resulting sequence of tokens.

\textbf{3) Embedding Layer:} The tokenized input \( t \) is converted into dense vector representations \( E \) which are optimized for model processing. \(
E = \text{Embedding}(t)
\), where \( t \) represents the sequence of tokens and \( E \) represents each token with a fixed dimensional embedding, of size \( d \).

\textbf{4) Attention Layer:} The proposed \textit{SmartVD} framework employs a Multi-Head Attention (MHA) mechanism within its attention layer. The self-attention process computes the attention scores \( A \) over the embeddings using the following equation:

\[
A = \text{Softmax}\left(\frac{Q \cdot K^T}{\sqrt{d_k}}\right) \cdot V
\]

where \( Q \), \( K \), and \( V \) represent the query, key, and value matrices derived from the embeddings \( E \), respectively. Here, \( d_k \) denotes the dimensionality of the key vectors. This mechanism enables the framework to effectively capture dependencies and relationships between tokens by computing weighted averages of the value vectors based on their relevance to the query vectors.

\textbf{6) LoRA Unit:} Low-Rank Adaptation (LoRA) is employed within specific layers of the Multi-Head Attention blocks to enhance the model’s capability for vulnerability prediction. LoRA introduces low-rank matrices that adjust the attention scores without significantly modifying the pre-trained weights. The modified attention scores \( A' \) are computed as:

\[
A' = A + \Delta A
\]

where \( \Delta A \) denotes the modifications introduced by LoRA. This approach enables efficient adaptation to vulnerability detection tasks with minimal computational overhead and parameter changes.

% These perturbations are derived from low-rank matrices that capture task-specific adaptations. By applying LoRA, the model effectively incorporates new task-related information while preserving the core knowledge embedded in the pre-trained parameters. This approach enables efficient adaptation to specialized tasks with minimal computational overhead and parameter changes.

\textbf{6) Normalization Layer: } Residual connections are incorporated into the outputs of both the self-attention and feed-forward layers, followed by normalization. The normalization process employs RMSNorm, which normalizes activations using the Root Mean Square (RMS) and scales them with learnable parameters. 

\[
\bar{a}_i = \frac{a_i}{\text{RMS}(a)} \cdot g_i, \quad \text{where} \quad \text{RMS}(a) = \sqrt{\frac{1}{n} \sum_{i=1}^{n} a_i^2}
\]
\newline \( a_i \) represents the pre-activation value for the \( i \)-th neuron, which is the sum of weighted inputs prior to applying the activation function. The term \( g_i \) is a learnable scaling parameter that adjusts the normalized output \( \bar{a}_i \), enabling the network to fine-tune the impact of each neuron's contribution. This technique ensures that the activations are normalized and scaled appropriately, improving the stability and performance of the model.

\textbf{6) Output Layer:} 
% In this layer of  \textit{smartVulDetect}, the framework produces text outputs \( O \) that provide predictions about vulnerabilities, including their types and severity.
In this component of the \textit{SmartVD} framework, the system generates text outputs \( O \) that deliver prediction regarding vulnerabilities, encompassing both their types and severity.
\[
O = \text{OutputLayer}(\text{Norm}_2)
\]

where \( \text{Norm}_2 \) is the output of the final normalization layer. The generated output includes information on the presence of vulnerabilities (Yes/No), the specific type of vulnerability (selected from 13 possible types), and the severity level (classified as Low, Medium, or High).

% \todo{More details about the output needs to be incorporated}
% The \texttt{OutputLayer} is a crucial component that maps the normalized representations to the target text space. For the Llama2 model, this typically involves a linear transformation followed by a softmax activation to produce a probability distribution over the vocabulary. Specifically:

% \[
% O = \text{Softmax}(W \cdot \text{Norm}_2 + b)
% \]

% where \( W \) denotes the weight matrix and \( b \) is the bias vector associated with the output layer. This mechanism enables the generation of coherent and contextually relevant text by transforming the model's internal state into meaningful output sequences. 

% \todo{Change the figure a bit}
\begin{figure*}[!htbp]
\centerline{\includegraphics[width=0.9\textwidth]{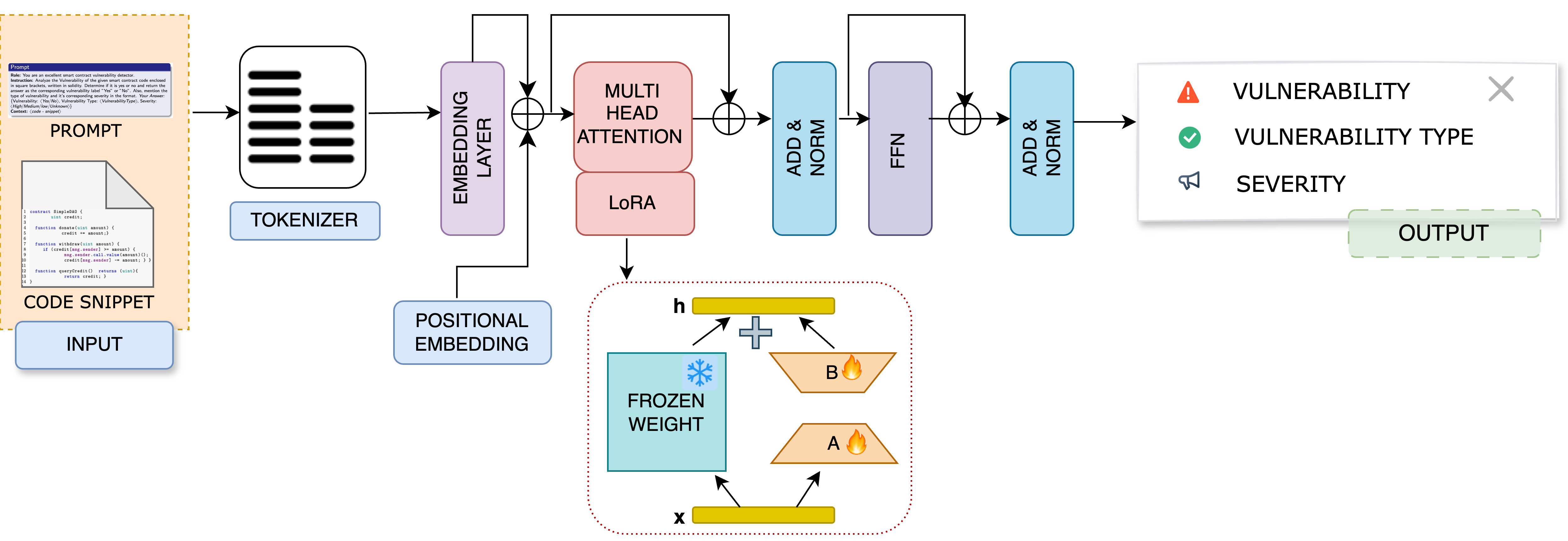}}
\caption{Proposed Model Architecture of Smart Contract Vulnerability Detection
Framework (SmartVD). A and B denote the LoRA modules. Frozen weight refers to the fixed weights of Multi-head attention. \(x\) and \(h\) are hidden representations before and after applying the LoRA module, respectively.}
\label{architecture}
\end{figure*}

\subsection{Alignment for Vulnerability detection: \textbf{Fine Tuning of Models}}

The framework incorporates LoRA adjustments, which fine-tune the model for specific tasks without substantial modification of the pre-trained weights. These adjustments are encapsulated in the learnable parameters of the attention layers, allowing for task-specific adaptations while preserving the core capabilities of the pre-trained model.

\textbf{GPT Models}
To evaluate the effectiveness of language models in smart contract vulnerability detection beyond open-source LLMs, we fine-tune GPT-3.5 Turbo\footnote{https://platform.openai.com/docs/models/gpt-3-5-turbo} and GPT-4o Mini\footnote{https://platform.openai.com/docs/models/gpt-4o-mini} for this specific task. Through prompt engineering, we generate diverse scenarios—zero-shot, few-shot, and step-by-step approaches—to guide the models in identifying vulnerabilities within Solidity smart contracts. The dataset is then converted into the JSON format required for GPT-3.5 Turbo and GPT-4o Mini fine-tuning, structuring each entry as a conversation with a system prompt, a user message containing the contract code, and an assistant response providing the expected output. Training and validation datasets are uploaded to OpenAI's servers, where fine-tuning jobs are initiated using the API, specifying the GPT-3.5 Turbo and GPT-4o Mini base models. This process effectively enhances the models' ability to accurately detect and classify vulnerabilities in smart contracts, making them more specialized for this novel task.

\textbf{Casual LLMs}
Given a dataset \( D = \{d_1, d_2, \dots, d_n\} \) where each \( d_i \) is a smart contract code snippet, we define the prompt function \( P \) such that:

\[
P(d_i) = \text{Prompt}(d_i) \implies \text{Input to the model}
\]
\newline For LoRA, the scaling factor \( \alpha \) is set to 16, the dropout rate \( \delta \) is 0.2, and the rank \( r \) is 4, with no bias \( \text{Bias} = 0 \). 
% The target modules \( T \) for adaptation are:
% % \[
% % T = \{\text{sel\_attn.q\_proj}, \, \text{self\_attn.k\_proj}, \, \text{self\_attn.v\_proj}, \, \text{self\_attn.o\_proj}, \, \text{mlp.gate\_proj}, \, \text{mlp.up\_proj}, \, \text{mlp.down\_proj}, \, \text{lm\_head}\}
% % \]
The target modules \( T \) are specifically adapted during the fine-tuning process. The training uses the optimizer \( \text{Optim} = \text{PagedAdamW-8bit} \), with a learning rate \( \eta = 1 \times 10^{-3} \) and a weight decay \( \lambda = 0.001 \). Mixed precision training is enabled using \( \text{fp16} \), and the maximum gradient norm is \( \text{max\_grad\_norm} = 0.3 \). The maximum number of training steps is \( \text{max\_steps} = 500 \), with a warm-up ratio \( \text{warmup\_ratio} = 0.03 \) and a cosine learning rate scheduler \( S(\eta, t) = \text{cosine} \). The evaluation strategy is based on steps, with logging steps set to \( \text{log\_steps} = 25 \). The total number of steps per epoch is calculated as:

\[
\text{Total Steps} = \left\lceil \frac{n}{B_{\text{device}} \times G} \right\rceil \times E
\]
where \( n \) is the total number of samples in the training dataset. The model parameters \( \theta \) are updated by minimizing the loss function \( \mathcal{L} \) using the specified optimizer:

\[
\theta_{\text{new}} = \theta_{\text{old}} - \eta \nabla_{\theta} \mathcal{L}(D, \theta)
\]

where \( D \) represents the training dataset. The process continues until the maximum number of steps \( \text{max\_steps} \) is reached. Following training, the fine-tuned model \( M_{\text{fine}} \) is merged to optimize deployment. The merged model \( M_{\text{merged}} \) is defined as:

\[
M_{\text{merged}} = \text{Merge}(M_{\text{base}}, \theta_{\text{fine}})
\]

\section{Experimental Results and Analysis}

\subsection{Experimental Setup}
We select six extensively used pre-trained large language models, comprising both code-related and general-purpose LLMs. We opt for two closed-source OpenAI models, GPT-3.5 Turbo and GPT-4o Mini, accessed through the OpenAI API Key. For open-source LLMs, we chose Llama2 (7B), CodeLlama (7B), CodeT5 (770M), and Falcon (7B), all accessible via the Hugging Face API\footnote{https://huggingface.co/}.
We fine-tune and execute all experiments involving open-source LLMs on a single-core NVIDIA A100 80GB PCIe server, utilizing its 80 GB GPU. For the fine-tuning and execution of experiments with closed-source LLMs, we leverage the OpenAI API.

\subsection{Qualitative Analysis}
% \todo{Use figures to support your answers}

% Initially, we selected few samples of smart contracts from test dataset and then used the standard prompt to perform queries for vulnerability detection along with its type and severity labels. As demonstrated in Figure \ref{}, we have passed smart contract solidity code as an input along with prompts , but the base model was not able to generate relevant outputs concernign the required vulnerabilty labels. However, when we fine tuned the model, as indicated in Figure \ref{}, and passed the same smart contract solidity code, our \textit{smartVD} framework was able to accurately detect Vulnerabilty in the solidity code along with the class to which the detected vulnerabilty  belongs to and the severity level of the vulneabilty class.
\begin{figure*}[!htbp]
    \subfloat[An instance of Codellama tested on \textit{VulSmart} dataset without fine-tuning\label{codellamaBase}]{%
      \includegraphics[width=0.45\textwidth]{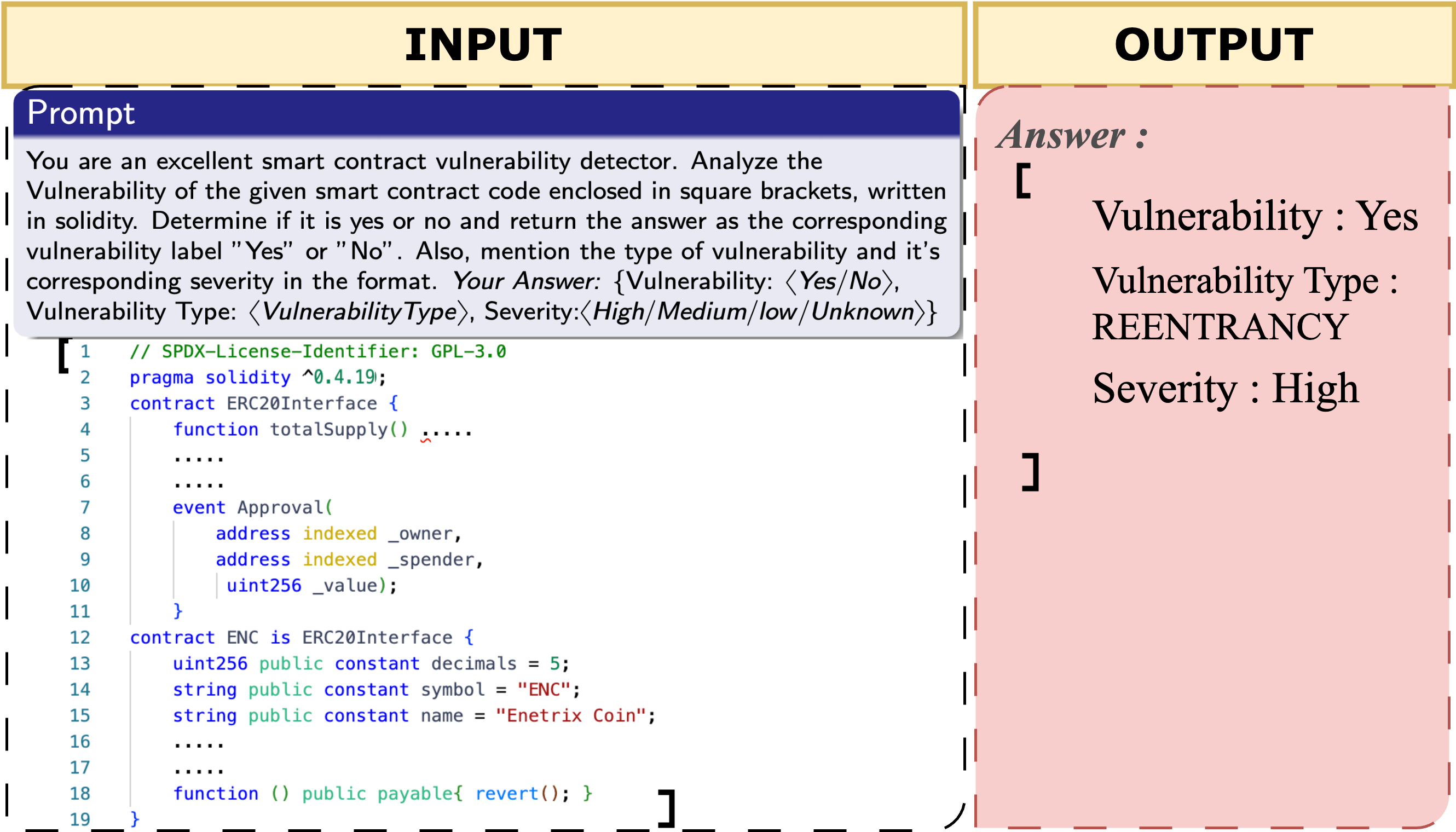}
    }
    \hfill
    \subfloat[An instance of Codellama tested on \textit{VulSmart} dataset after fine-tuning\label{codellamafine}]{%
      \includegraphics[width=0.45\textwidth]{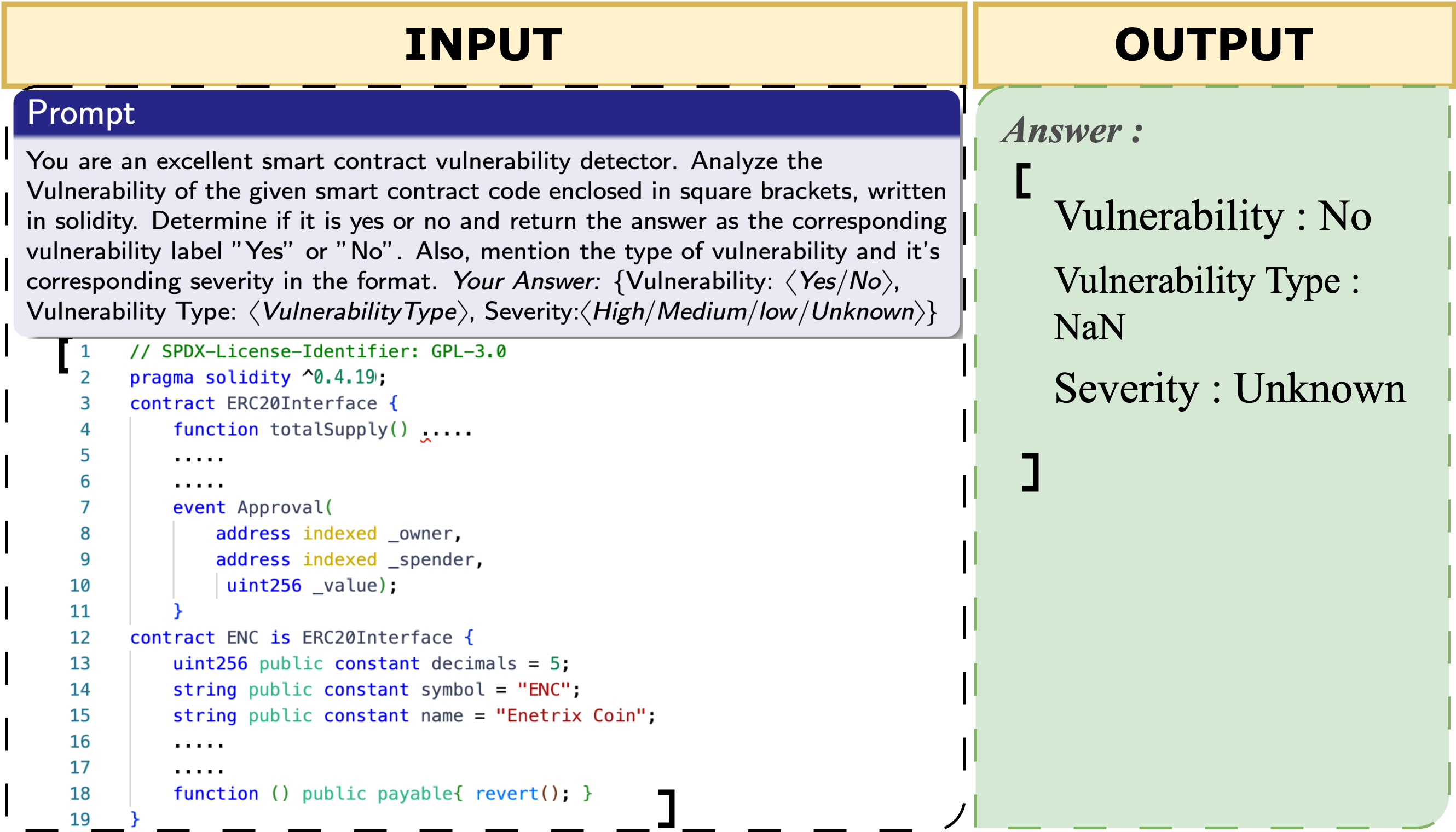}
    }\\
    \subfloat[An instance of Llama2 tested on \textit{VulSmart} dataset without fine-tuning\label{llama2base}]{%
      \includegraphics[width=0.45\textwidth]{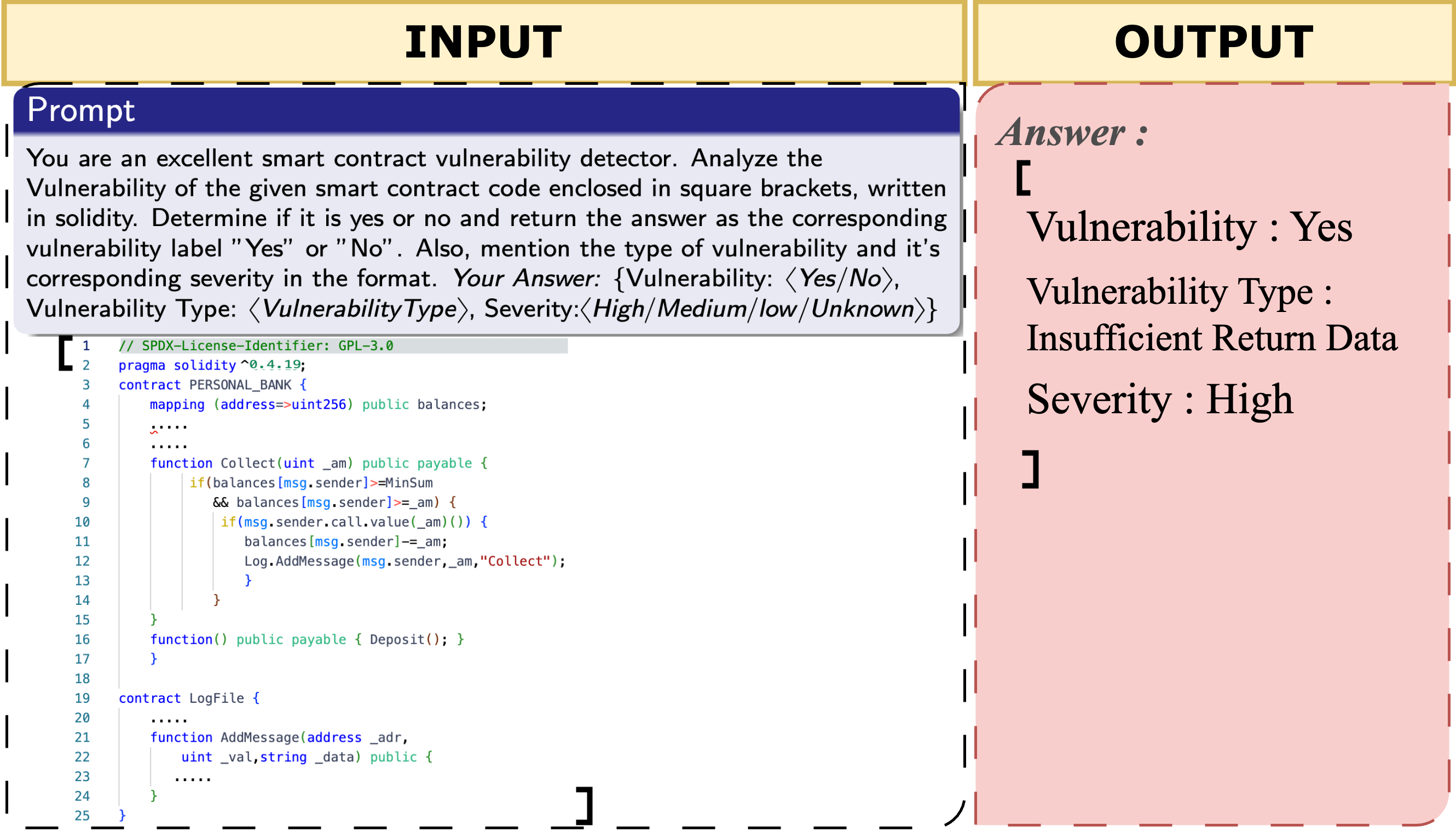}
    }
    \hfill
    \subfloat[An instance of Llama2 tested on \textit{VulSmart} dataset after fine-tuning\label{llama2fine}]{%
      \includegraphics[width=0.45\textwidth]{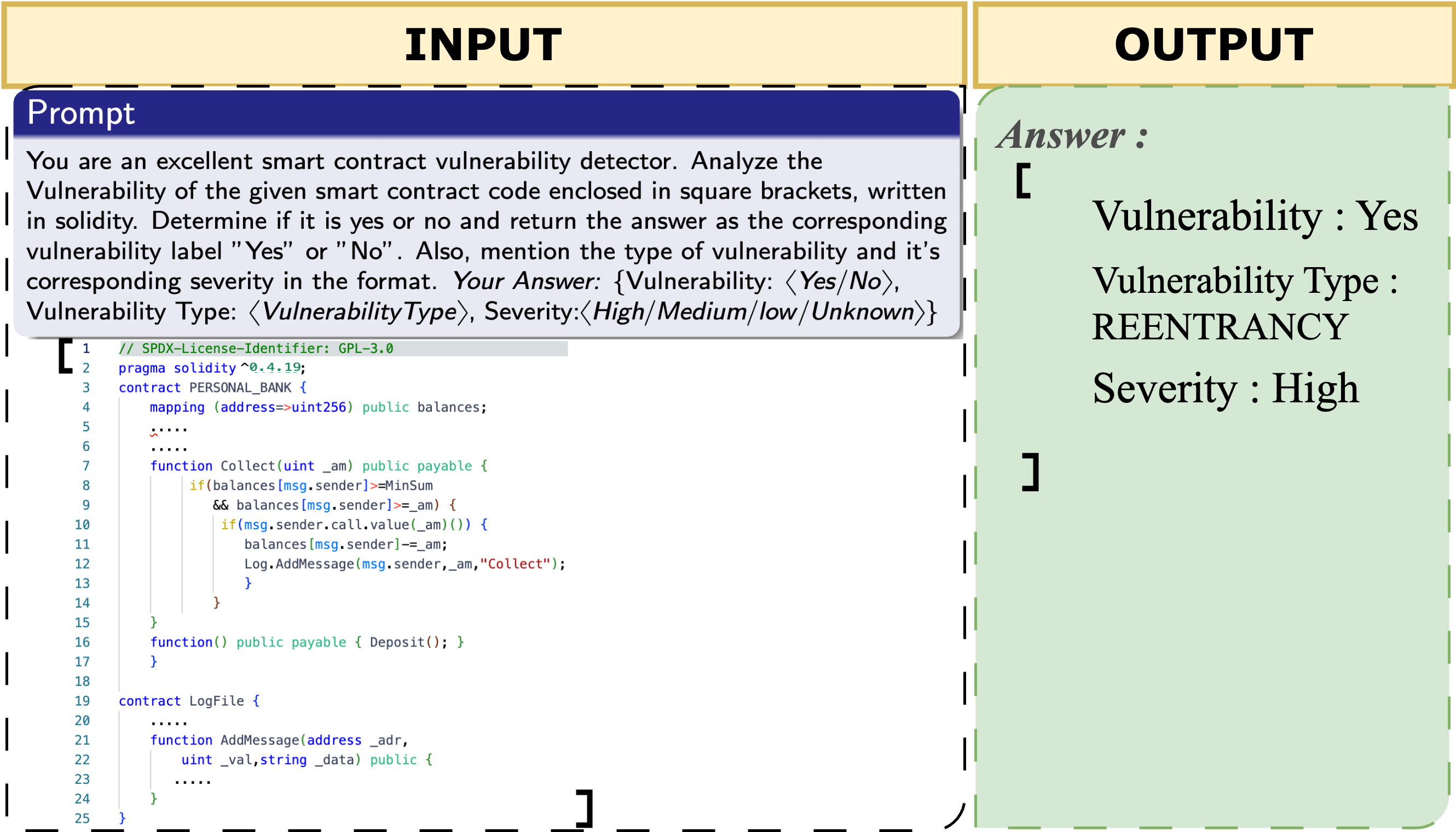}
    }
    \caption{Instance of LLMs on \textit{VulSmart} dataset}
    \label{fig:dummy}
  \end{figure*}

Initially, we select a few smart contract samples from the test dataset and use a standard prompt to query for vulnerabilities, including their types and severity levels. As shown in Figure \ref{codellamaBase}, we input the Solidity code along with the prompts, but the Codellama base model was unable to generate relevant outputs for the required vulnerability labels. However, after fine-tuning the model, as indicated in Figure \ref{codellamafine}, and using the same Solidity code, our \textit{SmartVD} framework accurately detect vulnerabilities, identify their corresponding classes, and determine the severity level of each vulnerability correctly. Similarly, as shown in Figure \ref{llama2base}, the base Llama2 model produces incoherent results when predicting the type of vulnerability. In contrast, the fine-tune version of Llama2 accurately identifies the vulnerability type, as illustrated in Figure \ref{llama2fine}. 

% \begin{figure}[!htbp]
% \centerline{\includegraphics[width=0.5\textwidth]{FiguresCompr/CodellamaBase.png}}
% \caption{An instance of Codellama tested on \textit{VulSmart} dataset without fine-tuning}
% \label{codellamaBase}
% \end{figure}

% \begin{figure}[!htbp]
% \centerline{\includegraphics[width=0.5\textwidth]{FiguresCompr/CodellamaFine.png}}
% \caption{An instance of Codellama tested on \textit{VulSmart} dataset after fine-tuning}
% \label{codellamafine}
% \end{figure}

% \begin{figure}[!htbp]
% \centerline{\includegraphics[width=0.5\textwidth]{FiguresCompr/Llama2Base.png}}
% \caption{An instance of Llama2 tested on \textit{VulSmart} dataset without fine-tuning}
% \label{llama2base}
% \end{figure}

% \begin{figure}[!htbp]
% \centerline{\includegraphics[width=0.5\textwidth]{FiguresCompr/Llama2Fine.png}}
% \caption{An instance of Llama2 tested on \textit{VulSmart} dataset after fine-tuning}
% \label{llama2fine}
% \end{figure}

\paragraph{Human Evaluation}

We select one PhD scholar and two post graduate students working in formal methods field and smart contract vulnerability detection to evaluate the performance of models. They used 10\% of the sample and gave a score out of 3 for correct vulnerability present, its class and severity. The average score obtained for our \textit{SmartVD} is 2.6 out of 3 (i.e. 87\% accuracy)

The observation here is that pre-trained model hallucinate and gives a large number of false positives in most cases. Whereas, the outputs generated by our \textit{SmartVD} framework is correct almost 87\% of the times in detecting the vulnerability with class and reduces the number of false positives.

\subsection{Quantitative Analysis}
% We have used Rogue, bleu ka descriptions

We have utilize four commonly employed evaluation metrics \textit{Accuracy, Precision, Recall,} and \textit{F1-score} to assess the performance of the models. Additionally we include BLEU score (Bilingual Evaluation Understudy) \cite{papineni2002bleu}, ROUGE score (Recall-Oriented Understudy for Gisting Evaluation) \cite{lin2005recall} and MCC score (Matthew's Correlation Coefficient) \cite{chicco2020advantages} to understand the change in improvement of model to solve the underlined problem statement. The BLEU score and ROGUE score are used for evaluating the quality of generated text. It measures the similarity between a machine-generated text and a reference text. The MCC metric help us to compare the same model with different configurations and accordingly rank transformer models. 

The MCC computes a measurement with true positives (TPs), true negatives (TNs), false positives (FPs), and false negatives (FNs).

The MCC can be summarized by the following equation:

\[
\text{MCC} = \frac{TP \times TN - FP \times FN}{\sqrt{(TP + FP)(TP + FN)(TN + FP)(TN + FN)}}
\]

The MCC value ranges from -1 to +1 where +1 means the prediction is 100\% accurate, 0 means model produces the random calculations and, -1 means the the predictions are totally false.
% and provides an excellent metric for classification models, even if the sizes of the classes are different. 

% We now have a good idea of how to measure a given transformer model’s results and compare them to other transformer models or NLP models.

% With some of the scoring methods in mind, we will now see how human evaluation can contribute to valuable measurements.
% % -\% increase,
% -best perfromimg model , etch etc,

% \subsection{Quantitative Analysis}
% We have used Rogue, bleu ka descriptions

% \todo{ have written a general performance measure across 6 tables.  have written in under the heading of Analysing the potential of LLMs for Smart Contract Vulnerabilities, with an aim to support the title of the paper.  can concise it, and then answer the research questions.  have not included here prompting techniques because that  guess will include while answering research question so did not wanted to do repeated information}

% \todo{LSTM,BiLSTM,GRU include multi-class classification results}

\paragraph{Analysing the potential of LLMs for Smart Contract Vulnerabilities:}
The performance of various language models (LLMs) on the \textit{VulSmart} dataset shows significant variations in their capability to classify vulnerabilities, vulnerability types, and severity levels. As shown in Table \ref{tab:table1}, we employ various LLMs for classifying vulnerabilities, vulnerability types, and severity levels. Among these LLMs, GPT-3.5 stands out as the top performer in vulnerability classification, achieving the highest accuracy (78\%), precision (82\%), recall (78\%), and F1-score (77\%). In comparison, CodeLlama which achieves a lower accuracy of 66\% and an F1-score of 59\%, indicating that while it performs well, it does not match the consistency and reliability of GPT-3.5. Additionally, traditional deep learning models like LSTM and GRU are utilized for the same classification tasks, but they struggle with accurate vulnerability classification, resulting in significantly lower precision, recall, and F1-scores. The only exception is BiLSTM, which exhibit significantly better performance in binary classification but shows a substantial decline in performance when applied to multi-class classification tasks.
% The significant difference between LLMs and traditional models demonstrates LLMs clear superiority in this domain.

\begin{table}[!hbtp]
\caption{Classification scores for vulnerability and its type on \textit{VulSmart} Dataset}
\label{tab:table1}
\begin{tabular}{ccccc|llll}
\hline
 & \multicolumn{4}{c|}{\textbf{Vulnerability (\%)}} & \multicolumn{4}{c}{\textbf{Vulnerability Type (\%)}} \\ \cline{2-9} 
 & \multicolumn{1}{c|}{\textit{Acc.}} & \multicolumn{1}{c|}{\textit{Pre.}} & \multicolumn{1}{c|}{\textit{Rec.}} & \textit{F1} & \multicolumn{1}{c|}{\textit{Acc.}} & \multicolumn{1}{c|}{\textit{Pre.}} & \multicolumn{1}{c|}{\textit{Rec.}} & \multicolumn{1}{c}{\textit{F1}} \\ \hline
\multicolumn{1}{c|}{LSTM} & 58 & 34 & 58 & 43 & 65 & 43 & 65 & 52 \\
\multicolumn{1}{c|}{BiLSTM} & 90 & 90 & 90 & 90 & 65 & 43 & 65 & 52 \\
\multicolumn{1}{c|}{GRU} & 58 & 34 & 58 & 43 & 65 & 43 & 65 & 52 \\
\multicolumn{1}{c|}{Llama2} & 56 & 43 & 56 & 43 & 64 & 41 & 64 & 50 \\
\multicolumn{1}{c|}{CodeLlama} & 66 & 74 & 66 & 59 & 64 & 41 & 64 & 50 \\
\multicolumn{1}{c|}{CodeT5} & 58 & 33 & 58 & 42 & 64 & 41 & 64 & 50 \\
\multicolumn{1}{c|}{Falcon} & 48 & 47 & 48 & 44 & 62 & 39 & 62 & 48 \\
\multicolumn{1}{c|}{GPT 3.5} & 78 & 82 & 78 & 77 & 52 & 46 & 52 & 48 \\
\multicolumn{1}{c|}{GPT-4o Mini} & 54 & 76 & 54 & 42 & 38 & 64 & 38 & 38 \\ \hline
\end{tabular}
\end{table}

Fine-tuning significantly boosts model performance, with GPT-3.5 showcasing the most consistent and significant improvements across all tasks. Open-source models, such as CodeLlama and Llama2, benefit greatly from fine-tuning, establishing them as attainable alternatives. As demonstrated in Table \ref{tab:table2}, GPT-3.5 shows the most remarkable improvement in vulnerability detection after fine-tuning, with an almost ideal MCC score of 1.000. Llama2 and CodeLlama additionally demonstrate significant progress, with MCC scores increasing from negative or low levels to 0.607831 and 0.717251, respectively. The progression continues in vulnerability type and severity label detection, with GPT-3.5 achieving an impressive MCC score of 0.897295 after fine-tuning. Similarly, CodeLlama and Llama2 exhibit significant improvements with fine-tuned MCC scores of 0.669070 and 0.599199, respectively, indicating improved generalization capabilities. However, CodeT5 shows no improvement after fine-tuning, with MCC scores close to zero, suggesting that it remains ineffective in vulnerability detection tasks despite fine-tuning.

\begin{table*}[!hbtp]
\centering
\caption{LLM performance improvement after finetune on \textit{VulSmart} Dataset}
\label{tab:table2}
\begin{tabular}{ccc|cc|cc}
\hline
\textbf{MODELS$\downarrow$} & \multicolumn{2}{c|}{\textbf{\begin{tabular}[c]{@{}c@{}}MCC\\ Score for \\ Vulnerability\end{tabular}}} & \multicolumn{2}{c|}{\textbf{\begin{tabular}[c]{@{}c@{}}MCC\\ Score for \\ Vulnerability Type\end{tabular}}} & \multicolumn{2}{c}{\textbf{\begin{tabular}[c]{@{}c@{}}MCC\\ Score for \\ Severity\end{tabular}}} \\ \cline{2-7} 
\textit{\textbf{}} & \multicolumn{1}{c|}{\textit{Base Model}} & \textit{\begin{tabular}[c]{@{}c@{}}Finetuned\\ Model\end{tabular}} & \multicolumn{1}{c|}{\textit{\begin{tabular}[c]{@{}c@{}}Base\\ Model\end{tabular}}} & \textit{\begin{tabular}[c]{@{}c@{}}Finetuned\\ Model\end{tabular}} & \multicolumn{1}{c|}{\textit{Base Model}} & \textit{\begin{tabular}[c]{@{}c@{}}Finetuned\\ Model\end{tabular}} \\ \hline
Llama2 & -0.048280 & 0.607831 & 0.082407 & 0.599199 & 0.265593 & 0.655856 \\
CodeLlama & 0.314945 & 0.717251 & 0.0 & 0.669070 & 0.223277 & 0.699974 \\
CodeT5 & 0.0 & 0.0 & 0.0 & 0.0 & -0.022811 & -0.056321 \\
Falcon & -0.048280 & 0.280900 & 0.082407 & 0.210609 & 0.265593 & 0.280908 \\
GPT 3.5 & 0.600245 & 0.999999 & 0.134973 & 0.897295 & 0.378698 & 0.967614 \\
GPT-4o Mini & 0.204124 & 0.999999 & 0.253056 & 0.900219 & 0.062896 & 0.934297 \\ \hline
\end{tabular}
\end{table*}

As indicated in Table \ref{tab:table3}, \ref{tab:table4}, and \ref{tab:table5} we compare the performance of base versus fine-tune models across various features, including vulnerability detection, vulnerability type, and severity level. Particularly, closed LLMs like GPT-3.5 and GPT-4o Mini exhibit the most significant improvements. For GPT-3.5, fine-tuning boosts accuracy from 78\% to 99\%, precision from 82\% to 99\%, and the F1-score from 77\% to 99\% in vulnerability detection tasks. CodeLlama also demonstrates substantial gains, particularly in precision and F1-score, which increase from 74\% and 59\% to 86\%, respectively. While Llama2 shows moderate improvement, CodeT5 does not benefit from fine-tuning, indicating its limitations for vulnerability detection. A comparable pattern is observed for vulnerability type and severity labels. Tables \ref{tab:table4} and \ref{tab:table5} provide more context into the models' performance in classifying vulnerability types and severity levels. GPT-3.5 and GPT-4o Mini continue to perform effectively, achieving near-perfect scores in fine-tuned models. In addition, CodeLlama showcases strong performance, particularly in the classification of vulnerability types, with considerable improvements in precision, recall, and F1-scores. However, models like CodeT5 and Falcon show only modest improvements, leaving GPT-3.5 and GPT-4o Mini as the most effective LLMs for the evaluated tasks, with fine-tuning greatly enhancing their capabilities.

\subsection{Observations}

\textbf{\textit{RQ1.}} \textit{How do various pre-trained LLMs and traditional deep learning techniques perform in detecting security vulnerabilities of Solidity smart contracts?}

The base models demonstrate considerable variation in performance across various metrics, indicating their varying underlying capabilities in detecting vulnerabilities. As demonstrated in Table \ref{tab:table1}, GPT-3.5 emerges as the best-performing base model, with 78\% accuracy, 82\% precision, 78\% recall, and a 77\% F1-score. This signifies that GPT-3.5 has a strong inherent ability to detect vulnerabilities, without additional training. CodeLlama additionally performs well as a base model, particularly in precision (74\%) and accuracy (66\%), revealing a relatively strong ability to identify vulnerabilities. Llama2 performs moderately, achieving 56\% in accuracy and recall and 43\% in precision and F1-score, which suggests that it can detect vulnerabilities but is less reliable than the top-performing models.
Falcon matches Llama2's moderate performance, with 48\% accuracy and recall, but slightly lower precision (47\%) and F1-score (44\%). CodeT5 base model is the model with the lowest precision (33\%), moderate accuracy (58\%), and recall (58\%). Its low F1-score (42\%) highlights its limited ability to detect vulnerabilities. Overall, there is a significant difference between the leading base models (GPT-3.5 and CodeLlama) and the lagging models (CodeT5, Falcon, and Llama2), highlighting that some models are better suited to vulnerability detection tasks in their pre-trained state. Additionally, deep learning techniques such as LSTM and GRU exhibit lower precision and F1-scores, achieving only 34\% and 43\%, respectively, compared to several pre-trained LLMs in vulnerability detection tasks. However, we observe improved accuracy, precision, and recall for the vulnerability detection task using BiLSTM. Despite this improvement, BiLSTM performance declines significantly when tasked with identifying the type of vulnerability in a multi-class classification scenario.

\textbf{\textit{RQ2.}}
\textit{How does fine-tuning of pre-trained LLMs using our proposed domain-specific dataset, \textbf{VulSmart}, impact the performance of vulnerability detection in smart contracts?}

\begin{table*}[!htbp]
\centering
\caption{LLM performance comparison for Vulnerability column on VulSmart Dataset}
\label{tab:table3}
\begin{tabular}{ccccccccc}
\hline
 & \multicolumn{8}{c}{\textbf{Vulnerability (Values in \%)}} \\ \cline{2-9} 
\textbf{MODELS$\downarrow$} & \multicolumn{2}{c|}{\textit{\textbf{Accuracy}}} & \multicolumn{2}{c|}{\textit{\textbf{Precision}}} & \multicolumn{2}{c|}{\textit{\textbf{Recall}}} & \multicolumn{2}{c}{\textit{\textbf{F1-score}}} \\ \cline{2-9} 
 & \multicolumn{1}{c|}{\textit{\begin{tabular}[c]{@{}c@{}}Base \\ Model\end{tabular}}} & \multicolumn{1}{c|}{\textit{\begin{tabular}[c]{@{}c@{}}Finetuned \\ Model\end{tabular}}} & \multicolumn{1}{c|}{\textit{\begin{tabular}[c]{@{}c@{}}Base\\ Model\end{tabular}}} & \multicolumn{1}{c|}{\textit{\begin{tabular}[c]{@{}c@{}}Finetuned \\ Model\end{tabular}}} & \multicolumn{1}{c|}{\textit{\begin{tabular}[c]{@{}c@{}}Base \\ Model\end{tabular}}} & \multicolumn{1}{c|}{\textit{\begin{tabular}[c]{@{}c@{}}Finetuned \\ Model\end{tabular}}} & \multicolumn{1}{c|}{\textit{\begin{tabular}[c]{@{}c@{}}Base \\ Model\end{tabular}}} & \textit{\begin{tabular}[c]{@{}c@{}}Finetuned \\ Model\end{tabular}} \\ \hline
Llama2 & 56 & \multicolumn{1}{c|}{62} & 43 & \multicolumn{1}{c|}{71} & 56 & \multicolumn{1}{c|}{62} & 43 & 61 \\
CodeLlama & 66 & \multicolumn{1}{c|}{86} & 74 & \multicolumn{1}{c|}{86} & 66 & \multicolumn{1}{c|}{85} & 59 & 86 \\
CodeT5 & 58 & \multicolumn{1}{c|}{58} & 33 & \multicolumn{1}{c|}{33} & 58 & \multicolumn{1}{c|}{58} & 42 & 42 \\
Falcon & 48 & \multicolumn{1}{c|}{64} & 47 & \multicolumn{1}{c|}{64} & 48 & \multicolumn{1}{c|}{64} & 44 & 64 \\
GPT 3.5 & 78 & \multicolumn{1}{c|}{99} & 82 & \multicolumn{1}{c|}{99} & 78 & \multicolumn{1}{c|}{99} & 77 & 99 \\
GPT-4o Mini & 54 & \multicolumn{1}{c|}{99} & 76 & \multicolumn{1}{c|}{99} & 54 & \multicolumn{1}{c|}{99} & 42 & 99 \\ \hline
\end{tabular}
\end{table*}

\begin{table*}[!htbp]
\centering
\caption{LLM performance comparison for Vulnerability Type column on VulSmart Dataset}
\label{tab:table4}
\begin{tabular}{ccccccccc}
\hline
 & \multicolumn{8}{c}{\textbf{Vulnerability Types (Values in \%)}} \\ \cline{2-9} 
\textbf{MODELS$\downarrow$} & \multicolumn{2}{c|}{\textit{\textbf{Accuracy}}} & \multicolumn{2}{c|}{\textit{\textbf{Precision}}} & \multicolumn{2}{c|}{\textit{\textbf{Recall}}} & \multicolumn{2}{c}{\textit{\textbf{F1-score}}} \\ \cline{2-9} 
 & \multicolumn{1}{c|}{\textit{\begin{tabular}[c]{@{}c@{}}Base \\ Model\end{tabular}}} & \multicolumn{1}{c|}{\textit{\begin{tabular}[c]{@{}c@{}}Finetuned \\ Model\end{tabular}}} & \multicolumn{1}{c|}{\textit{\begin{tabular}[c]{@{}c@{}}Base\\ Model\end{tabular}}} & \multicolumn{1}{c|}{\textit{\begin{tabular}[c]{@{}c@{}}Finetuned \\ Model\end{tabular}}} & \multicolumn{1}{c|}{\textit{\begin{tabular}[c]{@{}c@{}}Base \\ Model\end{tabular}}} & \multicolumn{1}{c|}{\textit{\begin{tabular}[c]{@{}c@{}}Finetuned \\ Model\end{tabular}}} & \multicolumn{1}{c|}{\textit{\begin{tabular}[c]{@{}c@{}}Base \\ Model\end{tabular}}} & \textit{\begin{tabular}[c]{@{}c@{}}Finetuned \\ Model\end{tabular}} \\ \hline
Llama2 & 64 & \multicolumn{1}{c|}{65} & 41 & \multicolumn{1}{c|}{86} & 64 & \multicolumn{1}{c|}{65} & 50 & 69 \\
CodeLlama & 64 & \multicolumn{1}{c|}{81} & 41 & \multicolumn{1}{c|}{81} & 64 & \multicolumn{1}{c|}{81} & 50 & 80 \\
CodeT5 & 64 & \multicolumn{1}{c|}{64} & 41 & \multicolumn{1}{c|}{41} & 64 & \multicolumn{1}{c|}{64} & 50 & 50 \\
Falcon & 62 & \multicolumn{1}{c|}{52} & 39 & \multicolumn{1}{c|}{52} & 62 & \multicolumn{1}{c|}{52} & 48 & 50 \\
GPT 3.5 & 52 & \multicolumn{1}{c|}{94} & 46 & \multicolumn{1}{c|}{93} & 52 & \multicolumn{1}{c|}{94} & 48 & 93 \\
GPT-4o Mini & 38 & \multicolumn{1}{c|}{94} & 64 & \multicolumn{1}{c|}{94} & 38 & \multicolumn{1}{c|}{94} & 38 & 94 \\ \hline
\end{tabular}
\end{table*}

\begin{table*}[!htbp]
\centering
\caption{LLM performance comparison for Severity column on VulSmart Dataset}
\label{tab:table5}
\begin{tabular}{ccccccccc}
\hline
 & \multicolumn{8}{c}{\textbf{Severity (Values in \%)}} \\ \cline{2-9} 
\textbf{MODELS$\downarrow$} & \multicolumn{2}{c|}{\textit{\textbf{Accuracy}}} & \multicolumn{2}{c|}{\textit{\textbf{Precision}}} & \multicolumn{2}{c|}{\textit{\textbf{Recall}}} & \multicolumn{2}{c}{\textit{\textbf{F1-score}}} \\ \cline{2-9} 
 & \multicolumn{1}{c|}{\textit{\begin{tabular}[c]{@{}c@{}}Base \\ Model\end{tabular}}} & \multicolumn{1}{c|}{\textit{\begin{tabular}[c]{@{}c@{}}Finetuned \\ Model\end{tabular}}} & \multicolumn{1}{c|}{\textit{\begin{tabular}[c]{@{}c@{}}Base\\ Model\end{tabular}}} & \multicolumn{1}{c|}{\textit{\begin{tabular}[c]{@{}c@{}}Finetuned \\ Model\end{tabular}}} & \multicolumn{1}{c|}{\textit{\begin{tabular}[c]{@{}c@{}}Base \\ Model\end{tabular}}} & \multicolumn{1}{c|}{\textit{\begin{tabular}[c]{@{}c@{}}Finetuned \\ Model\end{tabular}}} & \multicolumn{1}{c|}{\textit{\begin{tabular}[c]{@{}c@{}}Base \\ Model\end{tabular}}} & \textit{\begin{tabular}[c]{@{}c@{}}Finetuned \\ Model\end{tabular}} \\ \hline
Llama2 & 56 & \multicolumn{1}{c|}{64} & 41 & \multicolumn{1}{c|}{77} & 56 & \multicolumn{1}{c|}{64} & 43 & 63 \\
CodeLlama & 62 & \multicolumn{1}{c|}{83} & 54 & \multicolumn{1}{c|}{84} & 62 & \multicolumn{1}{c|}{83} & 53 & 83 \\
CodeT5 & 57 & \multicolumn{1}{c|}{55} & 33 & \multicolumn{1}{c|}{37} & 57 & \multicolumn{1}{c|}{55} & 42 & 42 \\
Falcon & 56 & \multicolumn{1}{c|}{58} & 61 & \multicolumn{1}{c|}{64} & 56 & \multicolumn{1}{c|}{58} & 45 & 55 \\
GPT 3.5 & 56 & \multicolumn{1}{c|}{98} & 69 & \multicolumn{1}{c|}{98} & 56 & \multicolumn{1}{c|}{98} & 57 & 98 \\
GPT-4o Mini & 34 & \multicolumn{1}{c|}{96} & 12 & \multicolumn{1}{c|}{96} & 34 & \multicolumn{1}{c|}{96} & 18 & 96 \\ \hline
\end{tabular}
\end{table*}

Fine-tuning significantly enhances the performance of various LLMs in detecting security vulnerabilities in smart contracts as proven by significant improvements in all key metrics, such as Accuracy, Precision, Recall, and F1-score as evident from Table \ref{tab:table3}, \ref{tab:table4}, and \ref{tab:table5}. GPT-3.5 and GPT-4o Mini emerge as the finest performers, with near-perfect scores of 99\% across all metrics after fine-tuning, confirming their brilliant suitability for vulnerability detection task. CodeLlama also shows strong performance, with its Precision, Recall, and F1-score all rising to 86\%, presenting it as a suitable framework for vulnerability detection.  On the other hand, CodeT5 remains static, showing no improvement across any metric after fine-tuning, indicating that it is less effective for this particular domain of vulnerability. Llama2 and Falcon show moderate improvements, but their scores remain lower than those of the leading models, indicating a limited ability to accurately detect vulnerabilities even after fine-tuning. Overall, fine-tuning is critical in improving the effectiveness of LLMs for detecting vulnerabilities in smart contracts, with some models benefiting significantly more than others.

To further evaluate the learning capabilities of the models, we assess their MCC scores, as presented in Table  \ref{tab:table2}. Among the six LLMs tested, all except CodeT5 show significant improvements in binary classification of vulnerability presence or absence. Llama2 led with the most impressive enhancement, boasting an MCC score increase of approximately 1359.19\%. Falcon came in second, with a notable MCC score improvement of around 682.00\%. Among closed-source models, GPT-4o Mini performed best, achieving an MCC score increase of about 389.96\%. CodeLlama and GPT 3.5 show improvements of roughly 127.74\% and 66.60\%, respectively.

We also explore the models' effectiveness in multi-class classification for detecting vulnerability types and their severity levels. All models demonstrate improvements in this area. CodeLlama excell in enhancing MCC scores for both vulnerability types and severity classes. Llama2 also shows significant progress, with percentage increases of approximately 627.14\% for vulnerability type and 147.00\% for severity class. The GPT models exhibit notable gains: GPT 3.5 Turbo has a percentage change of about 564.82\% for vulnerability type and 155\% for severity, while GPT-4o Mini mark increases of 255.65\% and 1385.83\% for type and severity, respectively. Falcon show moderate improvements, whereas CodeT5 remain relatively unchanged.

In summary, Llama2 stands out as the top-performing open-source model for binary classification, while GPT-4o Mini leads among closed-source models. For multi-class classification, CodeLlama outperforms Llama2, and GPT 3.5 surpasses GPT-4o Mini in learning capability.

We also note that GPT-4o Mini surpasses all other models in enhancing accuracy, precision, recall, and F1-score after fine-tuning on the \textit{VulSmart} dataset evident from Table \ref{tab:table3},\ref{tab:table4},\ref{tab:table5}.

\textbf{\textit{RQ3.}}
\textit{How do different prompting techniques influence the performance of both pre-trained and fine-tuned LLMs in vulnerability detection?}

% \usepackage{multirow}
\begin{table*}[!htbp]
\centering
\caption{Evaluation of various LLMs using `Zero Shot' Prompting strategy for the Multi Class Generative Framework}
\label{tab:table6}
\begin{tabular}{cccccccc}
\hline
 &  & \multicolumn{4}{c}{Metrics} &  &  \\ \cline{3-8} 
\textbf{Models $\downarrow$} &  & \textit{\textbf{BLEU-1}} & \textit{\textbf{BLEU-2}} & \textit{\textbf{BLEU-3}} & \textit{\textbf{ROUGE-1}} & \textit{\textbf{ROUGE-2}} & ROUGE-L \\ \hline
\multirow{2}{*}{\textit{Llama2}} & \textit{Base Model} & 0.613280 & 0.564826 & 0.553029 & 0.547381 & 0.44 & 0.547381 \\ \cline{2-2}
 & \textit{Finetuned Model} & 0.772603 & 0.728569 & 0.713511 & 0.742380 & 0.66 & 0.742380 \\ \hline
\multirow{2}{*}{\textit{CodeLLaMA}} & \textit{Base Model} & 0.412669 & 0.285044 & 0.256717 & 0.342619 & 0.17 & 0.342619 \\ \cline{2-2}
 & \textit{Finetuned Model} & 0.670691 & 0.575465 & 0.528571 & 0.640714 & 0.253704 & 0.640714 \\ \hline
\multirow{2}{*}{\textit{Code T5}} & \textit{Base Model} & 0.592543 & 0.523726 & 0.507637 & 0.542724 & 0.433333 & 0.542725 \\ \cline{2-2}
 & \textit{Finetuned Model} & 0.552835 & 0.474759 & 0.453883 & 0.512566 & 0.361111 & 0.512566 \\ \hline
\multirow{2}{*}{\textit{Falcon}} & \textit{Base Model} & 0.613280 & 0.564825 & 0.553028 & 0.547681 & 0.44 & 0.547681 \\ \cline{2-2}
 & \textit{Finetuned Model} & 0.619807 & 0.537561 & 0.517522 & 0.561905 & 0.44 & 0.561905 \\ \hline
\multirow{2}{*}{\textit{GPT 3.5}} & \textit{Base Model} & 0.613207 & 0.538773 & 0.505246 & 0.603334 & 0.38 & 0.603334 \\ \cline{2-2}
 & \textit{Finetuned Model} & 0.966539 & 0.954968 & 0.950576 & 0.965317 & 0.94 & 0.965317 \\ \hline
\multirow{2}{*}{\textit{GPT-4o Mini}} & \textit{Base Model} & 0.482868 & 0.365266 & 0.333199 & 0.404286 & 0.2 & 0.404286 \\ \cline{2-2}
 & \textit{Finetuned Model} & 0.968238 & 0.952963 & 0.948116 & 0.959603 & 0.94 & 0.959603 \\ \hline
\end{tabular}
\end{table*}

% \usepackage{multirow}
\begin{table*}[!hbtp]
\centering
\caption{Evaluation of various LLMs using `Few Shot' Prompting strategy for the Multi Class Generative Framework}
\label{tab:table7}
\begin{tabular}{cccccccc}
\hline
 &  & \multicolumn{4}{c}{Metrics} &  &  \\ \cline{3-8} 
\textbf{Models $\downarrow$} &  & \textit{\textbf{BLEU-1}} & \textit{\textbf{BLEU-2}} & \textit{\textbf{BLEU-3}} & \textit{\textbf{ROUGE-1}} & \textit{\textbf{ROUGE-2}} & ROUGE-L \\ \hline
\multirow{2}{*}{\textit{Llama2}} & \textit{Base Model} & 0.609764 & 0.537625 & 0.527099 & 0.549048 & 0.46 & 0.549048 \\ \cline{2-2}
 & \textit{Finetuned Model} & 0.815860 & 0.783319 & 0.771020 & 0.792222 & 0.7033 & 0.792222 \\ \hline
\multirow{2}{*}{\textit{CodeLLaMA}} & \textit{Base Model} & 0.468313 & 0.382972 & 0.356377 & 0.427857 & 0.22 & 0.427857 \\ \cline{2-2}
 & \textit{Finetuned Model} & 0.658608 & 0.559268 & 0.509802 & 0.617698 & 0.22 & 0.617698 \\ \hline
\multirow{2}{*}{\textit{Code T5}} & \textit{Base Model} & 0.562732 & 0.476467 & 0.460758 & 0.494761 & 0.38 & 0.494761 \\ \cline{2-2}
 & \textit{Finetuned Model} & 0.569017 & 0.480858 & 0.464543 & 0.504761 & 0.38 & 0.504761 \\ \hline
\multirow{2}{*}{\textit{Falcon}} & \textit{Base Model} & 0.633873 & 0.555114 & 0.539765 & 0.577143 & 0.46 & 0.577143 \\ \cline{2-2}
 & \textit{Finetuned Model} & 0.575797 & 0.515099 & 0.493571 & 0.559445 & 0.41 & 0.559445 \\ \hline
\multirow{2}{*}{\textit{GPT 3.5}} & \textit{Base Model} & 0.557467 & 0.473580 & 0.445105 & 0.504286 & 0.27 & 0.504286 \\ \cline{2-2}
 & \textit{Finetuned Model} & 0.908409 & 0.883564 & 0.877178 & 0.901905 & 0.86 & 0.901905 \\ \hline
\multirow{2}{*}{\textit{GPT-4o Mini}} & Base Model & 0.515509 & 0.403250 & 0.381018 & 0.450952 & 0.32 & 0.450952 \\ \cline{2-2}
 & Finetuned Model & 0.914166 & 0.898359 & 0.896202 & 0.904769 & 0.89 & 0.904769 \\ \hline
\end{tabular}
\end{table*}

\begin{table*}[!htbp]
\centering
\caption{Evaluation of various LLMs using `Chain of Thought' Prompting strategy for the Multi Class Generative Framework}
\label{tab:table8}
\begin{tabular}{cccccccc}
\hline
 &  & \multicolumn{4}{c}{Metrics} &  &  \\ \cline{3-8} 
\textbf{Models $\downarrow$} &  & \textit{\textbf{BLEU-1}} & \textit{\textbf{BLEU-2}} & \textit{\textbf{BLEU-3}} & \textit{\textbf{ROUGE-1}} & \textit{\textbf{ROUGE-2}} & ROUGE-L \\ \hline
\multirow{2}{*}{\textit{Llama2}} & \textit{Base Model} & 0.407071 & 0.255186 & 0.228060 & 0.317619 & 0.16 & 0.317619 \\ \cline{2-2}
 & \textit{Finetuned Model} & 0.763121 & 0.723618 & 0.714253 & 0.721746 & 0.65 & 0.721746 \\ \hline
\multirow{2}{*}{\textit{CodeLLaMA}} & \textit{Base Model} & 0.512181 & 0.417504 & 0.402112 & 0.447143 & 0.36 & 0.447143 \\ \cline{2-2}
 & \textit{Finetuned Model} & 0.692717 & 0.593354 & 0.540571 & 0.664365 & 0.23333 & 0.664365 \\ \hline
\multirow{2}{*}{\textit{Code T5}} & \textit{Base Model} & 0.620669 & 0.563046 & 0.554434 & 0.554047 & 0.5 & 0.554047 \\ \cline{2-2}
 & \textit{Finetuned Model} & 0.620669 & 0.563046 & 0.554434 & 0.554047 & 0.5 & 0.554047 \\ \hline
\multirow{2}{*}{\textit{Falcon}} & \textit{Base Model} & 0.407071 & 0.255186 & 0.228060 & 0.317619 & 0.16 & 0.317619 \\ \cline{2-2}
 & \textit{Finetuned Model} & 0.603370 & 0.541191 & 0.524359 & 0.551984 & 0.46 & 0.551984 \\ \hline
\multirow{2}{*}{\textit{GPT 3.5}} & \textit{Base Model} & 0.742481 & 0.685508 & 0.670385 & 0.701904 & 0.6 & 0.701904 \\ \cline{2-2}
 & \textit{Finetuned Model} & 0.771433 & 0.731304 & 0.710366 & 0.783095 & 0.63 & 0.783095 \\ \hline
\multirow{2}{*}{\textit{GPT-4o Mini}} & \textit{Base Model} & 0.623992 & 0.539558 & 0.518050 & 0.578217 & 0.47 & 0.578217 \\ \cline{2-2}
 & Finetuned Model & 0.941958 & 0.924206 & 0.918143 & 0.931269 & 0.9 & 0.931269 \\ \hline
\end{tabular}
\end{table*}

We evaluate the LLMs using various prompting techniques to assess any significant changes in their vulnerability detection capabilities. The performance of the generated outputs are measured against ground truth using BLEU and ROUGE metrics. We apply three widely-used prompting strategies (zero-shot, few-shot, and chain-of-thought) on each LLM. Detail descriptions of these prompts are provided in the appendix/supplementary section. We observe that base models and fine-tune models respond differently to each prompting technique.

Our analysis shows that the chain-of-thought prompting strategy yields the highest BLEU (BLEU-1, BLEU-2, BLEU-3) and ROUGE (ROUGE-1, ROGUE-2, ROUGUE-L) scores for the base models of CodeLlama, CodeT5, GPT-3.5, and GPT-4o Mini as evident from Table \ref{tab:table8}. For Llama2, zero-shot prompting proved most effective, while few-shot prompting worked best for Falcon's base model. Llama2’s lower performance with the chain-of-thought approach appears to result from the model being overly influenced by the examples and step descriptions provided in the prompt.

Post fine-tuning, we notice a shift in the models' performance across different prompting strategies. In general, the fine-tuned models perform better compared to their base models on the \textit{VulSmart} dataset, irrespective of the prompting techniques applied as evident from Table \ref{tab:table6}, \ref{tab:table7}, and \ref{tab:table8}. The only exception was Falcon, which did not show any improvement in few-shot prompting after fine-tuning. Specifically, the fine-tuned Llama2 model excels in few-shot prompting, while CodeLlama and CodeT5 perform best with chain-of-thought prompting. For zero-shot prompting, models like Falcon, GPT-3.5, and GPT-4o Mini deliver exceptional results.

% \textbf{CHANGE IN MCC SCORE for different prompting techniques.}

  %VULNERABILITY
  
% \begin{figure}[!hbtp]
%     \centering
%     \begin{tikzpicture}
%     \begin{axis}[
%         ybar=0.3cm,                   
%         bar width=0.15cm,             
%         width=0.45\textwidth,          
%         height=6cm,               
%          ymin=-0.06, ymax=1.000,      
%         ytick={-0.5,-0.09,0,0.2,0.4,0.6,0.8,1},     
%         ylabel={Value},              
%         symbolic x coords={Llama2, CodeLlama, CodeT5, Falcon, GPT-3.5 Turbo, GPT-4o Mini},
%         xtick=data,                   
%         nodes near coords,            
%         enlarge x limits=0.25,       
%         legend style={at={(0.5,-0.7)}, anchor=north,legend columns=-1,font=\small},
%         x tick label style={rotate=60, anchor=north, yshift=-0.3cm, xshift=-0.9cm}, 
%         xlabel={Models}
%     ]

%     % Base Model Data
%     \addplot+[ybar, color=violet, fill=violet!70] 
%         coordinates {
%             (Llama2, -0.05) (CodeLlama, 0.0) (CodeT5, 0.0) 
%             (Falcon, -0.05) (GPT-3.5 Turbo, 0.60) (GPT-4o Mini, 0.20)
%         };

%     % Fine-Tuned Model Data
%     \addplot+[ybar, color=teal, fill=teal!70] 
%         coordinates {
%             (Llama2, 0.61) (CodeLlama, 0.72) (CodeT5, 0.0) 
%             (Falcon, 0.28) (GPT-3.5 Turbo, 0.99) (GPT-4o Mini, 0.99)
%         };

%     \legend{Base Model, Fine-Tuned Model}

%     \end{axis}
% \end{tikzpicture}
%     \caption{MCC Scores for Vulnerability Classification using Zero-Shot across LLMs}
%     \label{MCC_zeroV}
% \end{figure}

\begin{figure}[!hbtp]
    \centering
    \begin{tikzpicture}
    \begin{axis}[
        ybar=0.3cm,                   
        bar width=0.15cm,              
        width=0.45\textwidth,         
        height=6cm,                  
        ymin=0, ymax=1.000,      
        ytick={-0.5,-0.09,0,0.2,0.4,0.6,0.8,1},     
        ylabel={Value},              
        symbolic x coords={Llama2, CodeLlama, CodeT5, Falcon, GPT-3.5 Turbo, GPT-4o Mini},
        xtick=data,                   
        nodes near coords,            
        enlarge x limits=0.25,        
        legend style={at={(0.5,-0.7)}, anchor=north,legend columns=-1,font=\small}, 
        x tick label style={rotate=60, anchor=north, yshift=-0.3cm, xshift=-0.9cm}, 
        xlabel={Models}
    ]

    % Base Model Data
    \addplot+[ybar, color=violet, fill=violet!70] 
        coordinates {
            (Llama2, 0.2) (CodeLlama, 0.09) (CodeT5, 0.1) 
            (Falcon, 0.28) (GPT-3.5 Turbo, 0.29) (GPT-4o Mini, 0.33)
        };

    % Fine-Tuned Model Data
    \addplot+[ybar, color=teal, fill=teal!70] 
        coordinates {
            (Llama2, 0.65) (CodeLlama, 0.65) (CodeT5, 0.14) 
            (Falcon, 0.45) (GPT-3.5 Turbo, 0.96) (GPT-4o Mini, 0.88)
        };

    \legend{Base Model, Fine-Tuned Model}

    \end{axis}
\end{tikzpicture}
    \caption{MCC Scores for Vulnerability Classification using Few-Shot across LLMs}
    \label{MCC_fewV}
\end{figure}

\begin{figure}[!hbtp]
    \centering
    \begin{tikzpicture}
    \begin{axis}[
        ybar=0.3cm,                   
        bar width=0.15cm,              
        width=0.45\textwidth,         
        height=6cm,                  
        ymin=0, ymax=1.000,      
        ytick={-0.5,-0.09,0,0.2,0.4,0.6,0.8,1},     
        ylabel={Value},              
        symbolic x coords={Llama2, CodeLlama, CodeT5, Falcon, GPT-3.5 Turbo, GPT-4o Mini},
        xtick=data,                   
        nodes near coords,            
        enlarge x limits=0.25,        
        legend style={at={(0.5,-0.7)}, anchor=north,legend columns=-1,font=\small}, 
        x tick label style={rotate=60, anchor=north, yshift=-0.3cm, xshift=-0.9cm}, 
        xlabel={Models}
    ]

    % Base Model Data
    \addplot+[ybar, color=violet, fill=violet!70] 
        coordinates {
            (Llama2, 0.11) (CodeLlama, -0.03) (CodeT5, -0.03) 
            (Falcon, 0.15) (GPT-3.5 Turbo, 0.29) (GPT-4o Mini, 0.23)
        };

    % Fine-Tuned Model Data
    \addplot+[ybar, color=teal, fill=teal!70] 
        coordinates {
            (Llama2, 0.66) (CodeLlama, 0.71) (CodeT5, 0.0) 
            (Falcon, 0.17) (GPT-3.5 Turbo, 0.76) (GPT-4o Mini, 0.83)
        };

    \legend{Base Model, Fine-Tuned Model}

    \end{axis}
\end{tikzpicture}
    \caption{MCC Scores for Vulnerability Types Class using Few-shot across LLMs}
    \label{MCC_fewVT}
\end{figure}

\begin{figure}[!hbtp]
    \centering
    \begin{tikzpicture}
    \begin{axis}[
        ybar=0.3cm,                   
        bar width=0.15cm,              
        width=0.45\textwidth,         
        height=6cm,                  
        ymin=0, ymax=1.000,      
        ytick={-0.5,-0.09,0,0.2,0.4,0.6,0.8,1},     
        ylabel={Value},              
        symbolic x coords={Llama2, CodeLlama, CodeT5, Falcon, GPT-3.5 Turbo, GPT-4o Mini},
        xtick=data,                   
        nodes near coords,            
        enlarge x limits=0.25,        
        legend style={at={(0.5,-0.7)}, anchor=north,legend columns=-1,font=\small}, 
        x tick label style={rotate=60, anchor=north, yshift=-0.3cm, xshift=-0.9cm}, 
        xlabel={Models}
    ]

    % Base Model Data
    \addplot+[ybar, color=violet, fill=violet!70] 
        coordinates {
            (Llama2, 0.16) (CodeLlama, 0.07) (CodeT5, 0.07) 
            (Falcon, 0.25) (GPT-3.5 Turbo, 0.25) (GPT-4o Mini, 0.29)
        };

    % Fine-Tuned Model Data
    \addplot+[ybar, color=teal, fill=teal!70] 
        coordinates {
            (Llama2, 0.69) (CodeLlama, 0.78) (CodeT5, 0.12) 
            (Falcon, 0.24) (GPT-3.5 Turbo, 0.80) (GPT-4o Mini, 0.80)
        };

    \legend{Base Model, Fine-Tuned Model}

    \end{axis}
\end{tikzpicture}
    \caption{MCC Scores for Severity Class using Few-shot across LLMs}
    \label{MCC_fewS}
\end{figure}

\begin{figure}[!hbtp]
    \centering
    \begin{tikzpicture}
    \begin{axis}[
        ybar=0.3cm,                   
        bar width=0.15cm,            
        width=0.45\textwidth,         
        height=6cm,                  
         ymin=0.0, ymax=1.000,      
        ytick={-0.5,-0.09,0,0.2,0.4,0.6,0.8,1},
        ylabel={Value},              
        symbolic x coords={Llama2, CodeLlama, CodeT5, Falcon, GPT-3.5 Turbo, GPT-4o Mini}, 
        xtick=data,                    
        nodes near coords,            
        enlarge x limits=0.25,       
        legend style={at={(0.5,-0.7)}, anchor=north,legend columns=-1,font=\small}, 
        x tick label style={rotate=60, anchor=north, yshift=-0.3cm, xshift=-0.9cm},
        xlabel={Models}
    ]

    % Base Model Data
    \addplot+[ybar, color=violet, fill=violet!70] 
        coordinates {
            (Llama2, 0.0) (CodeLlama, 0.12) (CodeT5, 0.25) 
            (Falcon, 0.0) (GPT-3.5 Turbo, 0.48) (GPT-4o Mini, 0.53)
        };

    % Fine-Tuned Model Data
    \addplot+[ybar, color=teal, fill=teal!70] 
        coordinates {
            (Llama2, 0.46) (CodeLlama, 0.72) (CodeT5, 0.25) 
            (Falcon, 0.25) (GPT-3.5 Turbo, 0.77) (GPT-4o Mini, 0.96)
        };

    \legend{Base Model, Fine-Tuned Model}

    \end{axis}
\end{tikzpicture}
    \caption{MCC Scores for Vulnerability Classification using Chain-of-thoughts across LLMs}
    \label{MCC_cotV}
\end{figure}

%VULNERABILITY TYPES

% \begin{figure}[!hbtp]
%     \centering
%     \begin{tikzpicture}
%     \begin{axis}[
%         ybar=0.3cm,                   
%         bar width=0.15cm,             
%         width=0.45\textwidth,          
%         height=6cm,               
%          ymin=-0.05, ymax=1.000,      
%         ytick={-0.5,-0.09,0,0.2,0.4,0.6,0.8,1},     
%         ylabel={Value},              
%         symbolic x coords={Llama2, CodeLlama, CodeT5, Falcon, GPT-3.5 Turbo, GPT-4o Mini},
%         xtick=data,                   
%         nodes near coords,            
%         enlarge x limits=0.25,       
%         legend style={at={(0.5,-0.7)}, anchor=north,legend columns=-1,font=\small},
%         x tick label style={rotate=60, anchor=north, yshift=-0.3cm, xshift=-0.9cm}, 
%         xlabel={Models}
%     ]

%     % Base Model Data
%     \addplot+[ybar, color=violet, fill=violet!70] 
%         coordinates {
%             (Llama2, 0.08) (CodeLlama, -0.01) (CodeT5, 0.0) 
%             (Falcon, 0.08) (GPT-3.5 Turbo, 0.13) (GPT-4o Mini, 0.25)
%         };

%     % Fine-Tuned Model Data
%     \addplot+[ybar, color=teal, fill=teal!70] 
%         coordinates {
%             (Llama2, 0.59) (CodeLlama, 0.67) (CodeT5, 0.0) 
%             (Falcon, 0.21) (GPT-3.5 Turbo, 0.89) (GPT-4o Mini, 0.90)
%         };

%     \legend{Base Model, Fine-Tuned Model}

%     \end{axis}
% \end{tikzpicture}
%     \caption{MCC Scores for Vulnerability Types Class using Zero-shot across LLMs}
%     \label{MCC_zeroVT}
% \end{figure}

\begin{figure}[!hbtp]
    \centering
    \begin{tikzpicture}
    \begin{axis}[
        ybar=0.3cm,                   
        bar width=0.15cm,            
        width=0.45\textwidth,         
        height=6cm,                  
         ymin=0.0, ymax=1.000,      
        ytick={-0.5,-0.09,0,0.2,0.4,0.6,0.8,1},
        ylabel={Value},              
        symbolic x coords={Llama2, CodeLlama, CodeT5, Falcon, GPT-3.5 Turbo, GPT-4o Mini}, 
        xtick=data,                    
        nodes near coords,            
        enlarge x limits=0.25,       
        legend style={at={(0.5,-0.7)}, anchor=north,legend columns=-1,font=\small}, 
        x tick label style={rotate=60, anchor=north, yshift=-0.3cm, xshift=-0.9cm},
        xlabel={Models}
    ]

    % Base Model Data
    \addplot+[ybar, color=violet, fill=violet!70] 
        coordinates {
            (Llama2, 0.0) (CodeLlama, 0.09) (CodeT5, 0.04) 
            (Falcon, 0.0) (GPT-3.5 Turbo, 0.52) (GPT-4o Mini, 0.4)
        };

    % Fine-Tuned Model Data
    \addplot+[ybar, color=teal, fill=teal!70] 
        coordinates {
            (Llama2, 0.54) (CodeLlama, 0.75) (CodeT5, 0.04) 
            (Falcon, 0.22) (GPT-3.5 Turbo, 0.37) (GPT-4o Mini, 0.87)
        };

    \legend{Base Model, Fine-Tuned Model}

    \end{axis}
\end{tikzpicture}
    \caption{MCC Scores for Vulnerability Types Class using Chain-of-thoughts across LLMs}
    \label{MCC_cotVT}
\end{figure}

%SEVERITY
% \begin{figure}[!hbtp]
%     \centering
%     \begin{tikzpicture}
%     \begin{axis}[
%         ybar=0.3cm,                   
%         bar width=0.15cm,             
%         width=0.45\textwidth,          
%         height=6cm,               
%          ymin=-0.08, ymax=1.000,      
%         ytick={-0.5,-0.09,0,0.2,0.4,0.6,0.8,1},     
%         ylabel={Value},              
%         symbolic x coords={Llama2, CodeLlama, CodeT5, Falcon, GPT-3.5 Turbo, GPT-4o Mini},
%         xtick=data,                   
%         nodes near coords,            
%         enlarge x limits=0.25,       
%         legend style={at={(0.5,-0.7)}, anchor=north,legend columns=-1,font=\small},
%         x tick label style={rotate=60, anchor=north, yshift=-0.3cm, xshift=-0.9cm}, 
%         xlabel={Models}
%     ]

%     % Base Model Data
%     \addplot+[ybar, color=violet, fill=violet!70] 
%         coordinates {
%             (Llama2, 0.27) (CodeLlama, 0.06) (CodeT5, -0.02) 
%             (Falcon, 0.27) (GPT-3.5 Turbo, 0.38) (GPT-4o Mini, 0.06)
%         };

%     % Fine-Tuned Model Data
%     \addplot+[ybar, color=teal, fill=teal!70] 
%         coordinates {
%             (Llama2, 0.66) (CodeLlama, 0.69) (CodeT5, -0.06) 
%             (Falcon, 0.28) (GPT-3.5 Turbo, 0.97) (GPT-4o Mini, 0.93)
%         };

%     \legend{Base Model, Fine-Tuned Model}

%     \end{axis}
% \end{tikzpicture}
%     \caption{MCC Scores for Severity Class using Zero-shot across LLMs}
%     \label{MCC_zeroS}
% \end{figure}

\begin{figure}[!hbtp]
    \centering
    \begin{tikzpicture}
    \begin{axis}[
        ybar=0.3cm,                   
        bar width=0.15cm,            
        width=0.45\textwidth,         
        height=6cm,                  
         ymin=0.0, ymax=1.000,      
        ytick={-0.5,-0.09,0,0.2,0.4,0.6,0.8,1},
        ylabel={Value},              
        symbolic x coords={Llama2, CodeLlama, CodeT5, Falcon, GPT-3.5 Turbo, GPT-4o Mini}, 
        xtick=data,                    
        nodes near coords,            
        enlarge x limits=0.25,       
        legend style={at={(0.5,-0.7)}, anchor=north,legend columns=-1,font=\small}, 
        x tick label style={rotate=60, anchor=north, yshift=-0.3cm, xshift=-0.9cm},
        xlabel={Models}
    ]

    % Base Model Data
    \addplot+[ybar, color=violet, fill=violet!70] 
        coordinates {
            (Llama2, 0.0) (CodeLlama, 0.07) (CodeT5, 0.11) 
            (Falcon, 0.0) (GPT-3.5 Turbo, 0.51) (GPT-4o Mini, 0.47)
        };

    % Fine-Tuned Model Data
    \addplot+[ybar, color=teal, fill=teal!70] 
        coordinates {
            (Llama2, 0.56) (CodeLlama, 0.90) (CodeT5, 0.11) 
            (Falcon, 0.21) (GPT-3.5 Turbo, 0.77) (GPT-4o Mini, 0.87)
        };

    \legend{Base Model, Fine-Tuned Model}

    \end{axis}
\end{tikzpicture}
    \caption{MCC Scores for Severity Class using Chain-of-thoughts across LLMs}
    \label{MCC_cotS}
\end{figure}
% \todo{A graph representing the difference in MCC score for different prompting technique. Each graph have both base and finetuned version.}

% \todo{Show the existing works in the field of LLMs lacks some capabilities that we have tackled in this paper.}
Table \ref{tab:table2} presents the Matthews Correlation Coefficient (MCC) scores, providing insight into the models' predictive accuracy before and after fine-tuning for both binary classification of vulnerability class and multi-class classification of vulnerability type and severity under zero-shot prompting. Fine-tuning notably improved the performance of all LLMs—open-source and closed-source—except for CodeT5, which maintained an MCC score of 0, indicating it generates random predictions rather than meaningful ones. In contrast, GPT models achieved an MCC score close to 0.99, demonstrating consistently accurate predictions.

The MCC scores for vulnerability detection, vulnerability type, and severity across all models using few-shot prompting are illustrated in Figures \ref{MCC_fewV}, \ref{MCC_fewVT}, and \ref{MCC_fewS}, respectively. Likewise, Figures \ref{MCC_cotV}, \ref{MCC_cotVT}, and \ref{MCC_cotS} display the MCC scores for the same tasks under chain-of-thought prompting. All models were fine-tuned using zero-shot prompting, yet we observe that fine-tuned CodeLlama outperformed other open-source LLMs across all prompting techniques, showing strong adaptability to the vulnerability detection tasks.

\textbf{\textit{RQ4:}} \textit{ To what extent do adversarial attacks affect the vulnerability detection capabilities of fine-tuned LLMs? }

Adversarial Attack is an attack where the goal is to cause an AI system to make a mistake or misclassification, often through subtle manipulations of the input data. To check how robust and good our models are to the adversarial attacks we took 50 non-vulnerable samples of solidity source codes. These 50 samples are detected non vulnerable by the LLMs implementing \textit{SmartVD} framework. We forcefully inject three class of bugs (namely Reentrancy, Arithmetic and tx.origin) to it using SolidiFi\cite{10.1145/3395363.3397385} tool.

\begin{table}[!htbp]

\caption{Fine-tuned LLM Performance Comparison Under Adversarial Attacks. }
\label{tab:table9}
\resizebox{0.49\textwidth}{!}{%
\begin{tabular}{c|ccc|ccc|ccc}
\hline
\multirow{2}{*}{\textbf{Models $\downarrow$}} & \multicolumn{3}{c|}{Vulnerability (\%)} & \multicolumn{3}{c|}{\textbf{Vulnerability Type (\%)}} & \multicolumn{3}{c}{\textbf{Severity (\%)}} \\ \cline{2-10} 
 & \multicolumn{1}{c|}{\textit{Pre.}} & \multicolumn{1}{c|}{\textit{Rec.}} & \textit{F1} & \multicolumn{1}{c|}{\textit{Pre.}} & \multicolumn{1}{c|}{\textit{Rec.}} & \textit{F1} & \multicolumn{1}{c|}{\textit{Pre.}} & \multicolumn{1}{c|}{\textit{Rec.}} & \textit{F1} \\ \hline
Llama2 & 100 & 22 & 36 & 44 & 22 & 30 & 72 & 22 & 34 \\
CodeLlama & 100 & 50 & 67 & 100 & 50 & 67 & 100 & 50 & 67 \\
CodeT5 & 0 & 0 & 0 & 0 & 0 & 0 & 0 & 0 & 0 \\
Falcon & 100 & 44 & 62 & 30 & 33 & 31 & 56 & 39 & 46 \\
GPT 3.5 & 100 & 100 & 100 & 100 & 100 & 100 & 100 & 100 & 100 \\
GPT-4o Mini & 100 & 100 & 100 & 100 & 100 & 100 & 100 & 100 & 100 \\ \hline
\end{tabular}
}
\end{table}

In our analysis, as shown in Table \ref{tab:table9} we found that among the open-source LLMs, the fine-tuned CodeLlama achieve a 100\% precision rate for detecting vulnerabilities, identifying their types, and assessing their severity. However, it record a recall of 50\% and an F1 score of 67\%, making it the best-performing model among the open-source LLMs considered for vulnerability detection. In contrast, the closed-source LLMs outperform the open-source models, achieving 100\% precision, recall, and F1 scores in detecting vulnerabilities, their types, and severity on an adversarial attacks dataset.

Interestingly, the recall and F1 scores of open-source LLMs decrease significantly when tested on adversarial smart contracts, indicating their limited effectiveness against such attacks. Conversely, the closed-source LLMs demonstrate complete resistance to adversarial attacks, with consistently perfect scores across all metrics, underscoring their reliability. This highlights the need for further research into improving the performance of open-source LLMs, potentially by incorporating more advanced semantic learning techniques alongside syntactic learning, to enhance their robustness against adversarial threats.    

% \todo{Report results in terms of recall, precision and f1 score}
% We observe that there is a decrease in performance on the LLMs  

% very bad performance on a new kind of attack

% is unable to identify semantics

% \todo{3.4 False Positive Analysis}
% refer to the paper page 11 of https://arxiv.org/pdf/2309.05520v4

% \textbf{\textit{RQ5:}} Impact of embedding on LLMs ie. AST, CFG, PDG. 
% \section{Conclusion}

% \section{Risk Analysis}
% While our \textit{SmartVD} model framework demonstrates promise, it is essential to have security experts and smart contract formal verification and auditing teams  to validate the findings, considering other critical factors. Our framework and dataset are intended to support security auditing professionals, not to replace their expertise.
% % rather than replace them.

\textit{\textbf{RQ5:} Can the integration of semantic structures, such as control-flow and data dependencies, enhance the performance of LLMs in identifying vulnerabilities?}

% \usepackage{multirow}
\begin{table}[!hbtp]
% \centering
\caption{Performance Comparison of GPT-4o Mini Across Various Semantic Structures}
\label{tab:table10}
\resizebox{0.49\textwidth}{!}{%
\begin{tabular}{llcccc}
\hline
 &  & \multicolumn{4}{c}{\textit{\textbf{Metrics (in \%)}}} \\ \cline{3-6} 
\textit{\textbf{Approach$\downarrow$}} &  & \textit{\textbf{Accuracy}} & \textit{\textbf{Precision}} & \textit{\textbf{Recall}} & \textit{\textbf{F1-score}} \\ \hline
\multirow{3}{*}{\textit{\textbf{Only Code}}} & \textit{Vulnerability} & 71 & 61 & 71 & 66 \\ \cline{2-2}
 & \textit{Vulnerability Type} & 24 & 39 & 24 & 23 \\ \cline{2-2}
 & \textit{Severity} & 41 & 22 & 41 & 28 \\ \hline
\multirow{3}{*}{\textit{\textbf{\begin{tabular}[c]{@{}l@{}}Code + \\ Control Flow \\ Graph (CFG)\end{tabular}}}} & \textit{Vulnerability} & 71 & 62 & 71 & 66 \\ \cline{2-2}
 & \textit{Vulnerability Type} & 24 & 39 & 24 & 23 \\ \cline{2-2}
 & \textit{Severity} & 44 & 37 & 44 & 40 \\ \hline
\multirow{3}{*}{\textit{\textbf{\begin{tabular}[c]{@{}l@{}}Code + Data \\ Dependency \\ information\end{tabular}}}} & \textit{Vulnerability} & 74 & 62 & 74 & 67 \\ \cline{2-2}
 & \textit{Vulnerability Type} & 26 & 43 & 26 & 24 \\ \cline{2-2}
 & \textit{Severity} & 44 & 47 & 44 & 37 \\ \hline
\multirow{3}{*}{\textit{\textbf{\begin{tabular}[c]{@{}l@{}}Code + CFG +\\ Data \\ Dependency\end{tabular}}}} & \textit{Vulnerability} & 79 & 81 & 79 & 80 \\ \cline{2-2}
 & \textit{Vulnerability Type} & 29 & 52 & 29 & 28 \\ \cline{2-2}
 & \textit{Severity} & 47 & 62 & 47 & 45 \\ \hline
\end{tabular}
}
\end{table}

To assess the impact of integrating semantic structures with Solidity code on the performance of large language models (LLMs), we augment the input prompt to the model with Control Flow Graph (CFG) and data dependency information. Specifically, we evaluate the performance of the GPT-4o Mini model by incorporating these additional semantic features. We select GPT-4o Mini for this evaluation due to its token limit capabilities, which allow for processing larger inputs without information loss. This is critical, as Solidity code combined with semantic structures like CFG and data dependencies often involves a significant number of tokens. With a maximum context window of 128k tokens, GPT-4o Mini is well-suited to handle these extensive inputs, ensuring comprehensive analysis during vulnerability detection in Solidity code. 

For our experiment, we select 50 smart contracts, both vulnerable and non-vulnerable, whose CFG and data dependency information could be extracted using the Slither tool \cite{feist2019slither}. Slither generates the control flow graph (CFG) for each function within a smart contract; however, it often encounters difficulties in generating complete CFGs or providing accurate data dependency information for many real-world smart contracts. The CFG information generated by Slither is in digraph dot language format, which we convert into a paired mapping representation, \texttt{CFG(Stmt1, Stmt2)}. This representation indicates that statement \texttt{Stmt2} is control-flow dependent on statement \texttt{Stmt1} for the given code statements (i.e. \texttt{Stmt2} is child of \texttt{Stmt1} within the control flow of the given code). The data dependency information for each function is provided in the form of key-value pairs, \texttt{DD(v,V)} where the \texttt{v} represents a unique variable, and the \texttt{V} is a list of variables on which \texttt{v} is dependent. To evaluate the effectiveness of GPT-4o Mini in detecting vulnerabilities, their types, and severity levels, we design four different categories of zero-shot prompts: (a) Solidity code alone, (b) Solidity code with its corresponding CFG, (c) Solidity code with its data dependency information, and (d) Solidity code with both CFG and data dependency information.

The performance of GPT-4o Mini under these various settings is summarized in Table \ref{tab:table10}, using metrics such as accuracy, precision, recall, and F1-score. Our analysis reveals that for binary classification of vulnerability presence, as well as multi-class classification of vulnerability type, incorporating CFG (Control Flow Graph) alongside the code in the prompts does not lead to noticeable improvements over using code alone. However, when it comes to multi-class classification of vulnerability severity, the inclusion of CFG with the code results in an accuracy increase from 41\% to 44\%, along with improvements in precision, recall, and F1-score compared to prompts using only the code. However, when data dependency information is added to the code, we observe a significant improvement in performance—accuracy increased from 71 to 74, precision increased from 61 to 62, and the F1-score improved from 66 to 67 for vulnerability class. Similarly, incorporating both CFG and data dependency information along with the code led to improved performance compared to using code, code with CFG and code with data dependency information.

% Similarly, incorporating both CFG and data dependency information with the code also resulted in improved performance compared to using only the code, though the combination of code and data dependency alone outperformed this configuration.

% Table \ref{tab:table10} refers to the performance of GPT-4o mini under various settings (i.e. Code, Code + CFG, Code + Data Dependency, Code + CFG + Data Dependency). We measure the performance through the metrics of accuracy, precision, recall and f1-score. We observe that the binary classification of presence of vulnerability, multi-class classification of vulnerability type and its severity for prompt with code and CFG is not good with lower accuracy precision and f1-score as compared to prompt with only code. But we see an improvement of precision from 63 to 76 and f1-score from 70 to 77 when code and data dependency information is passed as compared to only code information. We also see an improvement in metrics for vulnerability detection when code, CFG and data dependency information is applied as compared to only code information in prompt. However, Solidity code, CFG and data dependency stills performs less than code with data dependency information.

These results suggest that control flow information for each function does not contribute significant additional value to vulnerability detection. In contrast, leveraging data dependency information, both independently and in combination with control flow graph (CFG) information, significantly enhances the model's understanding of the underlying semantics, leading to better detection of vulnerabilities. The data dependency provides richer semantic insights for vulnerability detection compared to control-flow information. Additionally, we conclude that incorporating control flow information alongside data dependency significantly enhances model performance compared to using data dependency information alone.

\section{Threats to validity}

\textbf{External Validity:} The selection of large language models (LLMs) and datasets might introduce bias into our evaluation and findings. To minimize this risk, we utilize popular dataset corpus hosted on GitHub, containing
source code from real-world smart contracts that has
featured in peer-reviewed publications and create \textit{VulSmart} corpus. As the dataset labeling relies on formal methods-based tools, it may include false positives, which can adversely affect the training of the models.  Additionally, we include both state-of-the-art closed-source and open-source LLMs in our analysis. However, our insights may not be fully generalizable to other languages or datasets that are not covered in this study.

\textbf{Internal Validity:} Due to the inherent non-deterministic nature of LLMs and the fact that the test data is run only once, our observations can be influenced by this variability. To mitigate this, we set the temperature parameter near to zero for all LLMs to ensure determinism. While this method is effective for locally run open source models, it is known that GPT-4o Mini and GPT-3.5 might still produce non-deterministic results.

\textbf{Potential Bias in Evaluation:} Our evaluation scripts and code may contain bugs, which can introduce bias into the results. Furthermore, manual analysis of the outcomes can result in erroneous conclusions. To address these concerns, co-authors conduct regular reviews of the code, actively resolving any issues.

\section{Conclusion and future scope}

This paper presents a comprehensive approach to leveraging the power of large language models (LLMs) for detecting vulnerabilities in Solidity-based smart contracts. While much of the existing research highlights the potential of closed-source GPT models and Llama2, we introduce the \textit{Smart} Contract \textit{V}ulnerability \textit{D}etection Framework (\textit{SmartVD}), which demonstrates superior performance in vulnerability detection. Our findings indicate that among open-source LLMs, the fine-tuned CodeLlama (\textit{SmartVD}) outperforms and along with it perform better than closed-source base models GPT-3.5 Turbo and GPT-4o Mini. On the other hand, after fine-tuning, the closed-source GPT-3.5 Turbo and GPT-4o Mini models achieve the highest performance across both open and closed-source LLMs.

We conduct an in-depth analysis of LLMs on the \textit{VulSmart} dataset using zero-shot, few-shot, and chain-of-thought prompting techniques, with performance measured via BLEU and ROUGE scores. Additionally, we evaluate the robustness of these fine-tuned models under adversarial attack scenarios. Although their performance decrease under these conditions, the fine-tuned models still outperform their base versions. 

Although this work adapts multi-class classification based on the type of vulnerabilities present in the code and their severity levels, our future plan is to strengthen this by pinpointing the exact line or vulnerable function within the code. As highlighted in RQ5, incorporating data dependency in combination with control-flow information significantly improves the model's understanding of the underlying semantics, resulting in enhanced vulnerability detection. In future work, we plan to further explore the impact of integrating richer semantic information into LLM learning. This will include the use of Augmented Control Flow Graphs (CFG), Program Dependency Graphs (PDG), System Dependency Graphs (SDG), and pattern based code slicing information, alongside the source code. This integration has significant potential for advancing vulnerability detection tasks.

Notably, even in cases where the models misidentify the specific vulnerability type, they often correctly detect the presence of vulnerabilities. This suggests an exciting future direction: developing neuro-symbolic techniques that combine the intuitive reasoning abilities of LLMs with symbolic tools such as logical reasoning engines and static code analyzers. Such an approach could lead to more effective and interpretable solutions in smart contract vulnerability detection. 

%, a capability we plan to explore in future research.

\section*{Acknowledgments}
We graciously acknowledge the support of the Prime Minister Research Fellowship (PMRF) Award and the SERB Core Research Grant (Grant Number: CRG/2022/005794) by the Government of India for carrying out this research. We would like to express our gratitude to Soumyajit Mondal and Shobhit Gangawat from Mizoram University for their valuable assistance in conducting experiments during their internship at IIT Patna. 

\bibliographystyle{IEEEtran}
\bibliography{reference.bib}

\end{document}